\documentclass[%
 reprint,
 amsmath,amssymb,
 aps,
 prl,
]{revtex4-1}
\usepackage{enumitem}
\usepackage{graphicx}
\usepackage{dcolumn}
\usepackage{bm}

\usepackage[normalem]{ulem}

\usepackage[customcolors]{hf-tikz}  
\definecolor{myredcol}{RGB}{180,10,85} 

\usepackage{lipsum}
\usepackage{xfrac}
\usepackage{mathtools}

\usepackage[caption=false]{subfig} 
\usepackage{ragged2e} 
\DeclareCaptionJustification{justified}{\justifying}

\ExplSyntaxOff

\usepackage{pdfpages} 
\usepackage{pgffor} 
\usepackage{xr-hyper} 

\usepackage[unicode=true,pdfusetitle,
 bookmarks=true,bookmarksnumbered=false,bookmarksopen=false,
 breaklinks=true,pdfborder={0 0 0},pdfborderstyle={},backref=false,colorlinks=true]{hyperref}

\makeatletter
\AtBeginDocument{\let\LS@rot\@undefined}
\makeatother
\def\supplementfilename{supplement}
\pdfximage{\supplementfilename.pdf}
\def\numbersupplementpages{\the\pdflastximagepages}
\newif\ifarXiv
\arXivtrue 

\usepackage{xcolor,cancel}

\begin{document}

\preprint{APS/123-QED}

\title{Hydrodynamic description of Non-Equilibrium Radiation}

\author{Boris Rotstein}
\author{Eric Akkermans}%
\email{eric@physics.technion.ac.il}
\affiliation{%
Department of Physics, Technion -- Israel Institute of Technology, Haifa 3200003, Israel
}%

\date{\today}

\begin{abstract}

Non-equilibrium radiation is addressed theoretically by means of a stochastic lattice-gas model. We consider a resonating transmission line composed of a chain of radiation resonators, each at a local equilibrium, whose boundaries are in thermal contact with two blackbody reservoirs at different temperatures. In the long chain limit, the stationary state of the non-equilibrium radiation is obtained in a closed form. The corresponding spectral energy density departs from the Planck expression, yet it obeys a useful scaling form. A macroscopic fluctuating hydrodynamic limit is obtained leading to a Langevin equation whose transport parameters are calculated. In this macroscopic limit, we identify a local temperature which characterises the spectral energy density. The generality of our approach is discussed and applications for the interaction of non-equilibrium radiation with matter are suggested.

\end{abstract}

\pacs{Valid PACS appear here}
\maketitle

Radiation at thermal equilibrium has been a triggering problem underlying the quantum revolution \citep{planck1900theory,Einstein1917}. Since then, this problem has been revisited recurrently in different contexts either in physics or in engineering. Examples are abundant in condensed matter, statistical mechanics and quantum field theory among others \citep{glauber1963coherent,titulaer1966density,branczyk2017thermal,kindermann2002manipulation}. 

\begin{figure}[h]
\begin{centering}
\includegraphics[viewport=145bp 95bp 815bp 445bp,clip,width=1\columnwidth]{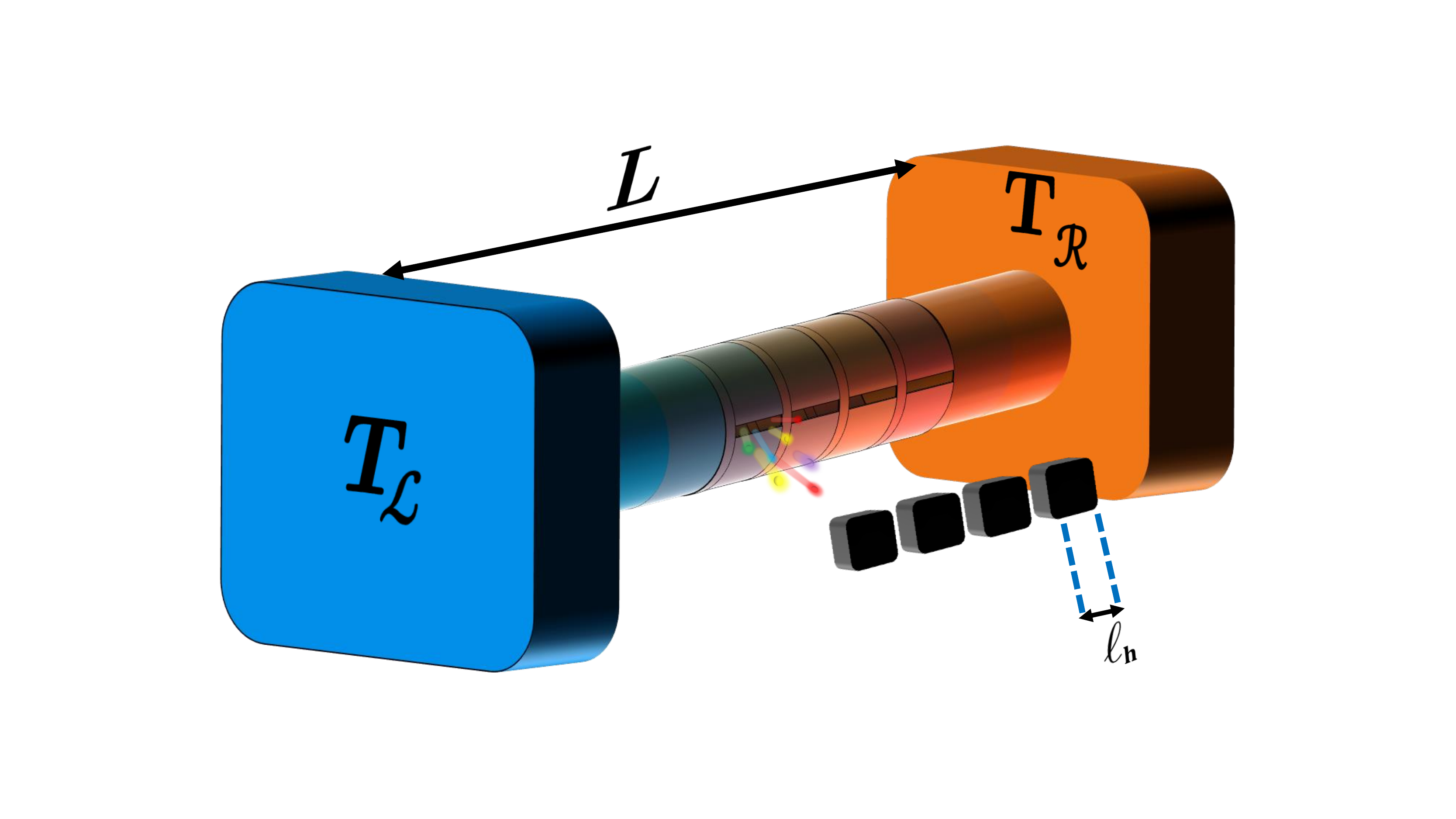}
\end{centering}
\caption{\label{fig:The-basic-setup} A resonating line connects two blackbody reservoirs held at distinct temperatures $T_{\mathcal{R}}\neq T_{\mathcal{L}}$. The spectral energy density $u_{\nu} (x)$ varies along the line  of length $L$. Adopting the standard measurement scheme of $u_{\nu} (x)$, the line is pierced at various locations and  an array of detectors of typical size $\ell_h$ (see text) records the local non equilibrium spectral radiation $u_{\nu} (x)$.}
\end{figure}

Radiation out of thermal equilibrium is a problem of wide and obvious interest, yet still largely uncharted despite important advances \citep{Chen2000,Joulain2005,Cleuren2007,Biehs2010,bunin2013transport,Nicacio2015,Greffet2018}. The purpose of this letter is to present a hydrodynamic description of non equilibrium radiation. While a wide range of problems can be formulated which involve non equilibrium radiation \citep{LEPRI20031,Bernard2012,Xuereb2015}, we focus to the case of two blackbody equilibrium radiation reservoirs held at different temperatures $T_{\mathcal{R}}\neq T_{\mathcal{L}}$ and connected by a properly designed long resonating line of length $L$ as sketched in Fig.\ref{fig:The-basic-setup}.  

Before dwelling into the details of our model, we now  summarize our main findings. 
Non equilibrium radiation is described using a coarse grained, boundary driven, microscopic lattice gas model for the energy transfer along the resonator, which accounts for hopping of photons between neighbouring cells of size $\ell$ $(\ell  \ll L)$. 
We show that this lattice gas model belongs to the well documented zero range process (ZRP) \citep{Evans2005}. The long time probability distribution $P_{\infty}\left(\eta\right)$ of photon configurations is obtained in (\ref{eq: product},\ref{eq: localeq}). Its continuous limit allows to identify a macroscopic
hydrodynamic regime for the steady state akin to the macroscopic fluctuation theory \cite{Bertini2015,Derrida2007}. In this regime, the fluctuating local spectral energy density $u_{\nu}\left(x,t\right)$ is constrained by a continuity equation,
\begin{equation}
    \partial_t u_{\nu}\left(x, t \right) = - \partial_x j(x,t)
    \label{continuity}
\end{equation}
where the fluctuating spectral current $j\left(x,t\right)$ obeys the Langevin equation,
\begin{equation}
j\left(x,t\right)=-D\left(\overline{u}_{\nu}\right)\partial_{x}u_{\nu}\left(x,t\right)+\sqrt{\sigma\left(\overline{u}_{\nu} \right)}\, \xi\left(x,t\right)\; .\label{eq:Langevin heat current}
\end{equation}
Here $\xi\left(x,t\right)$ is a weak $(L \gg 1)$ and delta-correlated white noise, 
$\overline{\xi\left(x,t\right)\xi\left(x',t'\right)}=\frac{1}{L}\delta\left(x-x'\right)\delta\left(t-t'\right)$. The transport coefficients, 
\begin{equation}
D\left(\overline{u}_{\nu}\right)=\frac{1}{\left(1+\overline{u}_{\nu}\right)^{2}}\, \, \text{and}\; \, \sigma\left(\overline{u}_{\nu}\right)=\frac{2\overline{u}_{\nu}}{1 +\overline{u}_{\nu}} \, , \label{eq:transp_coeff_via_SED}
\end{equation}
depend solely on the noise-averaged, local spectral energy density $\overline{u}_{\nu}\left(x\right)$ at frequency $\nu$,
\begin{equation}
\overline{u}_{\nu}\left(x\right) \equiv u_{\nu} \left( x \right)/g_{\nu} h \nu \,   = \frac{z_{\mathcal{L}}+\frac{x}{L}\left(z_{\mathcal{R}}-z_{\mathcal{L}}\right)}{1-z_{\mathcal{L}}-\frac{x}{L}\left(z_{\mathcal{R}}-z_{\mathcal{L}}\right)}\; ,\label{sed}
\end{equation}
where $z_{\mathcal{R}/\mathcal{L}} \equiv \exp\left(-h\nu/k_{B}T_{\mathcal{\mathcal{R}}/\mathcal{L}}\right)$  and $g_{\nu} \equiv 8 \pi \nu^2 / c^3$ is the $3d$ density of states of the radiation. This expression, very distinct from the Planck distribution, abides the scaling (\ref{eq: scalingform}) which allows to identify a macroscopic local temperature function $T_\tau (x)$ at the hydrodynamic scale. A measurement of local radiation fluxes through apertures along the transmission line akin to standard blackbody measurements is displayed in Fig.\ref{fig:The-basic-setup}. It  provides an experimental way to directly access $u_{\nu}\left(x\right)$ and probe the predicted scaling form and its departure from Planck distribution.

The rest of the letter is devoted to a description of our model and setup, to a derivation of the results just stated and finally, to a discussion of their meaning and applicability.

\begin{figure}[h]
\begin{centering}
\includegraphics[viewport=0bp 195bp 960bp 545bp,clip,width=1\columnwidth]{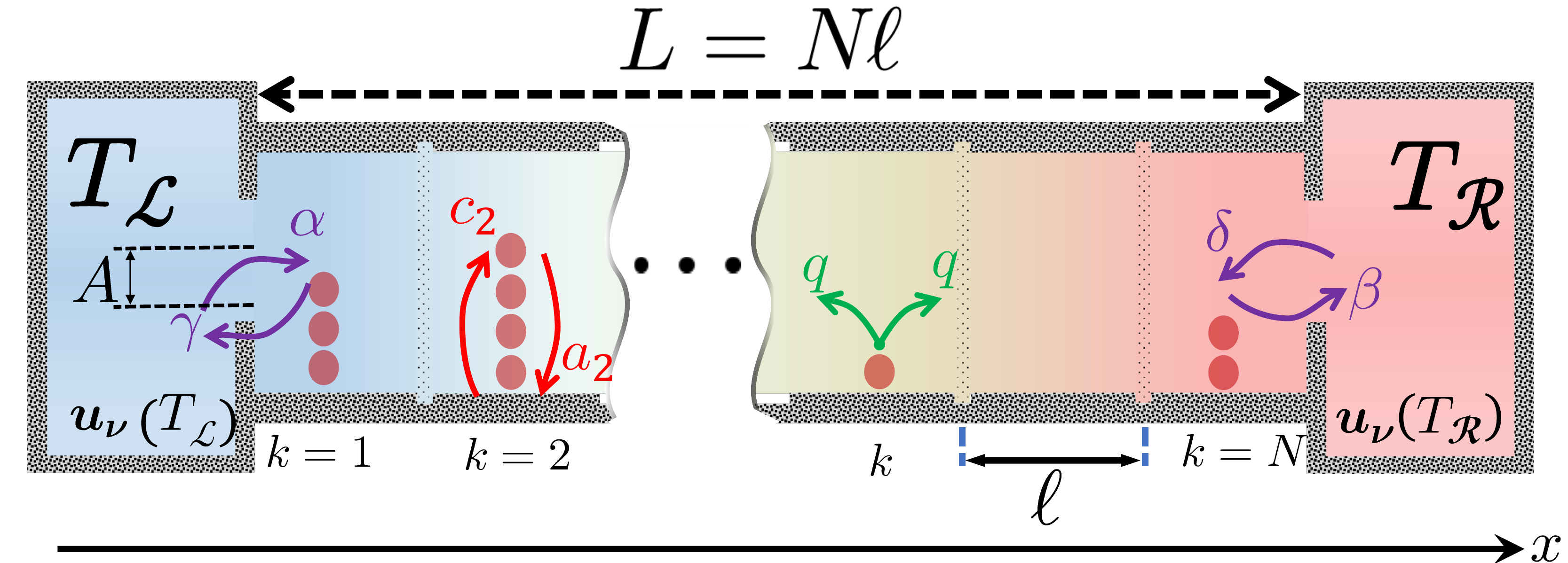}
\end{centering}
\caption{\label{fig:1D The-lattice-model} 
Examples of photon configurations in the lattice model with corresponding  transition
rates. Each cell $k$ of size $\ell$ accounts for a lattice site, where radiation is locally at thermal equilibrium with surrounding walls. The three  processes (1,2,3) namely, in-cell creation and annihilation (red), bulk (green) and boundaries (purple) inter-cell exchanges are indicated. 
}
\end{figure}

\emph{The Physical Setup \textendash{}} The two blackbody radiation reservoirs in Fig.\ref{fig:The-basic-setup} are held at distinct temperatures $T_{\mathcal{R}}\neq T_{\mathcal{L}}$ and are respectively characterised by the Planck distributions,  $u_{\nu}\left(T_{r}\right)=g_{\nu}h\nu \left(\mathrm{e}^{\sfrac{h\nu}{k_{B}T_{r}}}-1\right)^{-1},\quad {r} =\mathcal{L},\mathcal{R}$ of their spectral energy densities at frequency $\nu$. They are connected by a long transmission line of length $L$ built out of a series of resonators, hereafter  cells. In this setup\textcolor{myredcol}{,} illustrated in Fig.\ref{fig:1D The-lattice-model}, we assume that each cell $k$ of size $\ell$, is large enough so as the enclosed radiation is at thermal equilibrium. 
For $L \gg \ell$, the density of cells is 
finite and the precise nature of the coupling between neighboring resonators is unimportant. Local thermal equilibrium with the walls (the environment) is achieved  through local Kirchhoff law, a point appropriately explored in \citep{Greffet2018}. 

\emph{The Model \textendash{}}  We now outline the assumptions underlying the microscopic lattice model used for the description of the radiation in the transmission line. We consider a one-dimensional
lattice $k = 1, \dots, N$ of $N = L / \ell$ cells. The left and right blackbody reservoirs are respectively located at $k=0$ and $k=N+1$, so that $T_{\mathcal{L}}\equiv T_{0}$ and $T_{\mathcal{R}}\equiv T_{N+1}$. The radiation consists of a gas of photons occupying the lattice cells and hopping between neighboring cells. This hopping is  described by random transitions between cells occupation numbers configurations $\eta=\left\{ n_{1},\dots,n_{N}\right\} $, with $n_{k}=0,1,2,\ldots$, i.e $n_{k}$ is unbounded. An example of a configuration is displayed in Fig.\ref{fig:1D The-lattice-model}.

We denote  $\eta_{\overline{i},\underline{j}}\equiv \left\{ n_{1},\dots,n_{i}+1,\ldots,n_{j}-1,\ldots,n_{N}\right\}$, $\quad n_{j}>0$, a photon configuration which relatively to $\eta$, has an excess of one photon in the $i^{\mathrm{th}}$ cell and a depletion of one photon in cell $j$. For the bulk and boundary cells we consider the three following processes:
\begin{enumerate}
\item \emph{Creation/annihilation}: $\eta\xrightleftharpoons[a_{k}\left(n_{k}+1\right)]{c_{k}\left(n_{k}\right)}\eta_{\overline{k}}$ photons are absorbed at the boundaries of the $k^{\text{th}}$ cell at a rate $a_{k}$ and created at a rate $c_{k}$. These rates depend on $n_k$ and we set $a_{k} =0$ for $n_k = 0$ for all $k$. Furthermore, these rates are constrained by local detailed balance (\ref{detailedbalance}) within each cell $k$.

\item \emph{Bulk exchange}: $\eta_{\overline{k},k+1}\xrightleftharpoons[q]{q}\eta_{k,\overline{k+1}}$, namely the hopping of photons  between neighboring cells is symmetric and occurs at a  rate $q$. We shall set $q=1$ in the sequel. 
\item \emph{Boundary processes \textendash{}} At the boundary cells  $k=1$ and $k=N$, we assign: 
\begin{itemize}
    \item[i)] $\eta\xrightleftharpoons[\gamma]{\alpha}\eta_{\overline{1}}$ : The left reservoir respectively absorbs and injects photons from/into the transmission line at rates $\gamma$ and $\alpha$.
    
    \item[ii)] $ \eta\xrightleftharpoons[\beta]{\delta}\eta_{\overline{N}}$ : The right reservoir respectively absorbs and injects photons from/into the transmission line at rates $\beta$ and $\delta$.
\end{itemize} 
\end{enumerate}
Physically, the rates $\alpha$ and $\delta$ express the radiation fluxes from the
reservoir into the transmission line through small apertures of area $A$, namely, $\alpha=2A\,c\,u_{\nu}\left(T_0\right)/h$ and  $\delta=2A\,c\,u_{\nu}\left(T_{N+1}\right)/h$ (see Fig.\ref{fig:1D The-lattice-model} and Supplemental Material S\,I \citep{suppmat1}).

The three processes $(1,2,3)$ listed above, underlie the dynamics of non equilibrium radiation states by means of the probability $P_{t}\left(\eta\right)$ for a configuration $\eta$ of photons at time $t$. This dynamics of configurations is thus given by solutions of a master equation, $\partial_{t}P_{t}\left(\eta\right)=\mathcal{L}\left[P_{t}\left(\eta\right)\right]$ where the generator $\mathcal{L}$, linear in $P_{t}$, specifies the rate of  probability flow between occupation configurations. We are interested in the  form of the  long time probability
$P_{\infty}\left(\eta\right)\equiv\lim_{t\rightarrow\infty}P_{t}\left(\eta\right)$, namely in the solutions of the kernel equation,
\begin{equation}
\mathcal{L}\left[P_{\infty}\left(\eta\right)\right] = 0 \, .    
\label{nerkernel}
\end{equation}
Since the  processes $(1)$ and $(2+3)$ are independent, we look for solutions with independent generators,
\begin{equation}
    \mathcal{L} \equiv \mathcal{L}^{\perp} + \mathcal{L}^{\parallel}
    \label{parperp}
\end{equation}
where $\mathcal{L}^{\perp}$ accounts for process $(1)$ and $\mathcal{L}^{\parallel} \equiv \mathcal{L}_{\text{bulk}}+\mathcal{L}_{\text{boundary}}$ accounts for processes $(2)$ and $(3)$. Expressions  of these generators are given in the Supplementary Material S\,IV in \citep{suppmat1}.
To evaluate the kernel of $\mathcal{L}^{\parallel}$,
we consider the statistics of the total current $Q_t $ of independent photons hopping on the lattice,  
and removed and injected from the boundary blackbody reservoirs at rates ($\alpha, \beta, \gamma, \delta$) during the time interval $[0,t]$. 
We define the cumulant generating function 
$\langle \mathrm{e}^{\lambda Q_t} \rangle $ of $Q_t$, the average $\langle \cdots \rangle$ being on the Poisson processes governing the time-dependent boundary dynamics (S\,II in \citep{suppmat1}). In the long time limit, we expect $\langle \mathrm{e}^{\lambda Q_t} \rangle \simeq \mathrm{e}^{\mu^{\parallel} (\lambda) t}$.
Hence, the knowledge of $\mu^{\parallel} (\lambda )$ allows to obtain the cumulants of $Q_t$ by
\begin{equation}
    \lim_{t \rightarrow \infty} \frac{\langle Q_t ^n \rangle_c}{t} = \frac{\mathrm{d}^n \mu (\lambda )}{\mathrm{d} \lambda^n}|_{\lambda = 0} \, .
\end{equation}
To compute $\mu^{\parallel} (\lambda )$, we rely on the assumption that photons are independent\textcolor{myredcol}{,} which allows to study separately the effect of right and left reservoirs.  
The number of photons leaving the transmission line from its left boundary in the time interval $[0,t]$ is, 
\begin{equation}
    Q^{\mathcal{L}}_{[0,t]} = \overrightarrow{{\bf 1}}_{[0,t]} - \overleftarrow{{\bf 1}}_{[0,t]} 
    \label{leftcurrent}
\end{equation}
where $\overrightarrow{{\bf 1}}_{[0,t]}$ (resp. $\overleftarrow{{\bf 1}}_{[0,t]}$) is the number of photons leaving (resp. entering) the left reservoir to (resp. from) cell ${\bf 1}$  in the time interval $[0,t]$. Since photons are independent, we decompose the total current by partitioning the time interval $[0,t]$ into $N_t$ segments $[t_k,t]$ with $t_k$ being the time at which the $k^{\text{th}}$ photon entered the system. The total current $Q^{\mathcal{L}}_{[0,t]}$ then appears as the sum of elementary contributions of each photon that has entered the resonator between times $0$ and $t$, $Q^{\mathcal{L}}_{[0,t]} = \sum_{k =0}^{N_t} Q^k _{[t_k,t]}$. This partition simplifies the calculation of the left part of the cumulant generating function $\mu_\mathcal{L} (\lambda) = \lim_{t \rightarrow \infty} \frac{1}{t} \ln \langle e^{\lambda  Q^{\mathcal{L}}_{[0,t]}} \rangle$ by factorising it into a product of single photon contributions (see S\,II in \citep{suppmat1} for details). A similar calculation for the  current $Q^{\mathcal{R}}_{[0,t]}$ of photons leaving the transmission line from its right boundary and the corresponding generating function $\mu_\mathcal{R} (\lambda)$ leads to \citep{suppmat1},
\begin{equation}
    \mu^{\parallel} (\lambda) = \mu_\mathcal{L}  - \mu_\mathcal{R} 
    = \frac{\alpha \beta ( e^\lambda -1 ) - \gamma \delta ( 1 - e^{- \lambda} )}{\gamma ( \beta N +1 ) - \beta (\gamma -1)} \, .
    \label{mu}
\end{equation}
The generating function $ \mu^{\parallel} (\lambda)$ coincides with those describing  the dynamics of a class of stochastic lattice gas models known as the zero-range-process (ZRP) \citep{Harris2005}. This result is not obvious since, unlike our model, the ZRP describes interacting particles. Yet, based on this identity, we use the result, proven for ZRP \citep{Levine2005}, that the long time probability $P_{\infty}\left(\eta\right)$ solution of $\mathcal{L}^{\parallel} \left[P_{\infty}\left(\eta\right)\right] = 0$ in (\ref{nerkernel}), is a product measure, namely,
\begin{equation}
    P_{\infty}\left(\eta\right)=\prod_{k=1}^{N}\pi_{k}\left(n_{k}\right) \, .
    \label{eq: product}
\end{equation}
Each term $\pi_{k}\left(n_{k}\right)$ accounts for the bookkeeping of photon occupation number in cell $k$ at local equilibrium, and it is expressed in terms of the steady state fugacities $z_{k}$,
\begin{equation}
    \pi_{k}\left(n_{k}\right)=\left(1-z_{k}\right)z_{k}^{n_{k}} \, .
    \label{eq: localeq}
\end{equation}
Under this form, a sufficient condition for local equilibrium is expressed by a  detailed balance condition,
\begin{equation}
    \frac{c_{k}\left(n_{k}\right)}{a_{k}\left(n_{k}+1\right)}=  \frac{P_{\infty}\left(\eta_{\overline{k}}\right)}{P_{\infty}\left(\eta\right)} = z_{k}  \, .
    \label{detailedbalance}
\end{equation}
Expressions (\ref{eq: localeq}) and (\ref{detailedbalance}) generalize the condition for thermal blackbody radiation with fugacity $z ^{B} = \exp\left(-{h\nu}/{k_{B}T }\right)$ \citep{suppmat1}. 
It is immediate to check that (\ref{eq: localeq}) implies 
$ \mathcal{L}^{\perp}\left[P_{\infty}\left(\eta\right)\right] =0 $. Hence (\ref{nerkernel}) amounts to solutions of $\mathcal{L}^{\parallel}\left[P_{\infty}\left(\eta\right)\right]= \mathcal{L}_{\text{ZRP}} \left[P_{\infty}\left(\eta\right)\right] = 0$ characterised by fugacities \citep{Levine2005,suppmat1}, 
 \begin{equation}
z_{k}=\frac{\frac{k}{N}\left(\gamma\delta-\alpha\beta\right)+\alpha\beta-\frac{1}{N}\left[\gamma\delta+ \left(\alpha+\delta\right)\right]}{\beta\gamma\left(1-\frac{1}{N}\right)+\frac{1}{N}\left(\beta+\gamma\right)} \, .
\label{zk}
\end{equation}
Fugacities in the blackbody reservoirs are given by $z_{\mathcal{L}/ \mathcal{R} } \equiv z_{{0}/{N+1}} =  \exp\left(-h\nu/k_{B}T_{\mathcal{L}/ \mathcal{R}} \right)$. Taking the large $N$ limit in (\ref{zk}) leads to the boundary conditions \citep{Derrida2007},
\begin{equation}
   z_{0} = \mathrm{e}^{-\frac{h\nu}{k_{B}T_{\mathcal{L}}}}=  \frac{\alpha}{\gamma}  ,\qquad z_{N+1} = \mathrm{e}^{-\frac{h\nu}{k_{B}T_{\mathcal{R}}}} = \frac{\delta}{\beta}\;, \label{boundarycond} 
\end{equation}
\footnote{Having $0<z_{0/L+1}<1$ imposes $\gamma>\alpha$ and $\beta>\delta$.}, so that (\ref{mu}) rewrites,
\begin{equation}
  \mu^{\parallel} (\lambda) =  \frac{1}{L} \left( 1 - \mathrm{e}^{-\lambda} \right) \left(z_{\mathcal{L}} \, \mathrm{e}^\lambda - z_{\mathcal{R}} \right) \, .
  \label{muperp}
\end{equation}
Boundary conditions (\ref{boundarycond}) can also be obtained in a different way if one notes that $\mu^{\parallel} (\lambda)$ in (\ref{muperp}) abides the Gallavotti and Cohen relation \citep{Gallavotti1995,Lebowitz1998},
\begin{equation}
 \mu^{\parallel} (\lambda) = \mu^{\parallel} (- \lambda -E)
 \label{GC}
\end{equation}
where $E$ is a field that brings the radiation out of equilibrium. Taking $E \equiv \ln z_{\mathcal{R}} - \ln z_{\mathcal{L}} $, corresponds to (\ref{boundarycond}).

To establish (\ref{sed}) for the spectral energy density $u_{\nu}\left(x\right)$, we now consider the hydrodynamic continuous limit obtained by averaging over cell sizes $\ell$. Namely\textcolor{myredcol}{,} defining $k\ell/L = k / N \equiv x \; \left(0\leq x \leq 1\right)$,  with $L\rightarrow\infty$, $\ell\rightarrow\infty$, and keeping a finite density of cells $\ell/L\rightarrow\text{d} x$. This averaging procedure , applied to the fugacity in \eqref{zk} gives,
\begin{equation}
 z_{k}\rightarrow z\left(x\right)=\frac{\alpha}{\gamma}+x\left(\frac{\delta}{\beta}-\frac{\alpha}{\gamma}\right) = z_{\mathcal{L}}+x\left(z_{\mathcal{R}}-z_{\mathcal{L}}\right)  \, .
\label{continuum1}
\end{equation}
The spectral energy density of the radiation at frequency $\nu$ in cell $k$ inside the transmission line is  $u_{\nu}\left(k\right) = g_{\nu} h \nu \left\langle n_k \right\rangle$ with $\left\langle n_k \right\rangle = \sum_{\eta}n_{k}P_{\infty}\left(\eta\right) = z_k / ( 1 - z_k )$. In the continuous limit, $\left\langle n\left(x\right) \right\rangle=\frac{z\left(x\right)}{1-z\left(x\right)}$ leads to (\ref{sed}) for the macroscopic spectral energy density $u_{\nu}\left(x\right)$ as announced in the introductory part. 
\begin{center}
\begin{figure}[h]
\begin{centering}
\subfloat[\label{fig: 2graphs(a)}]{
    \includegraphics[viewport=0bp 10bp 880bp 600bp,clip,width=0.5\columnwidth]{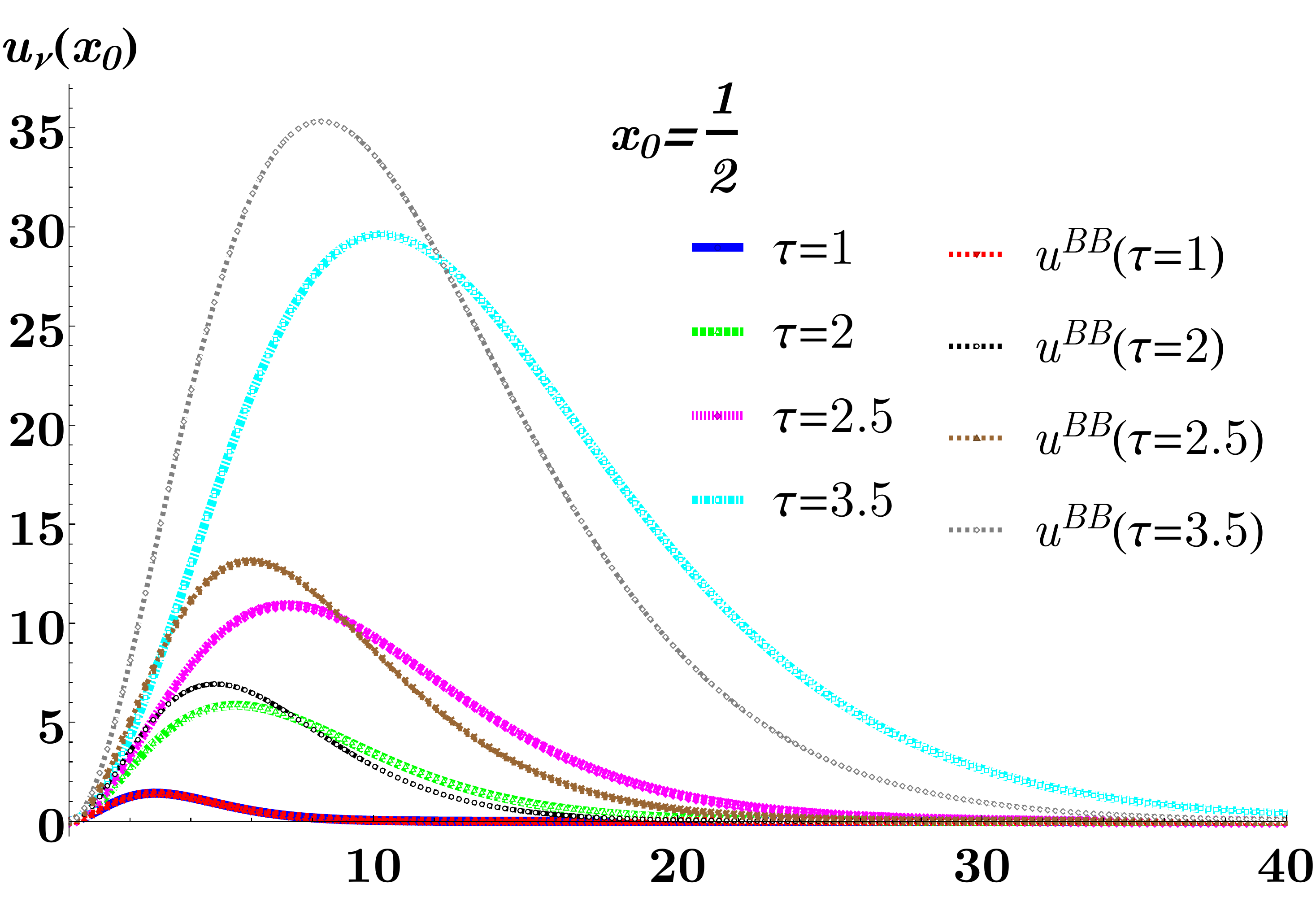}
    }
\subfloat[\label{fig: 2graphs(b)}]{
   \hspace{-1em}\includegraphics[viewport=0bp 20bp 920bp 590bp,clip,width=0.5\columnwidth]{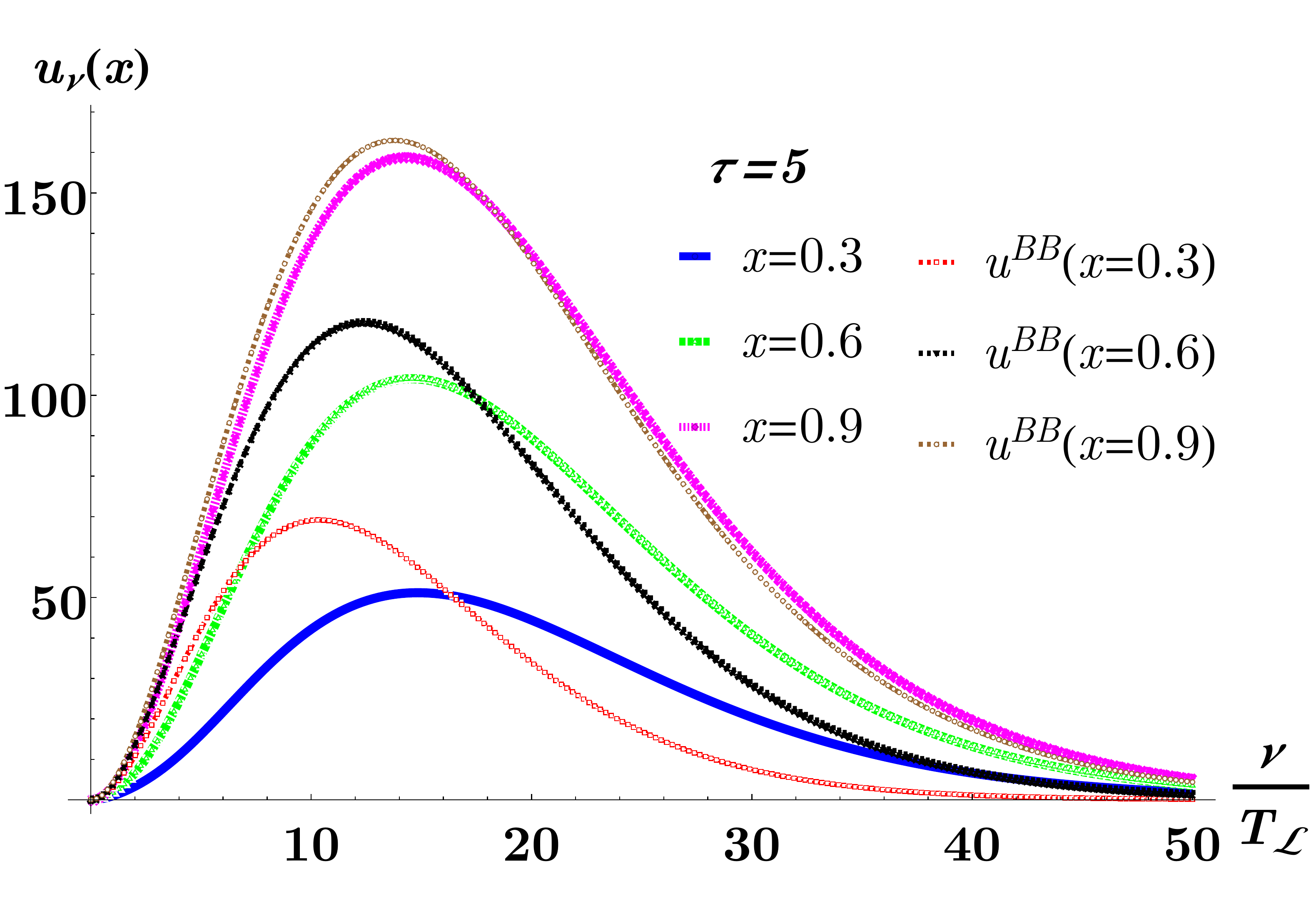}    }
\end{centering}
\caption{\label{fig: 2graphs} Behaviour of $u_{\nu}\left(x \right)$ in (\ref{sed}) - Difference with the Planck spectral distribution. (a) $u_{\nu}\left(x = 1/2 \right)$, plotted as a function of $\nu / T_{\mathcal{L}}$ for different values of the ratio $\tau \equiv T_{\mathcal{R}} / T_{\mathcal{L}}$. The Planck distribution $u^{B}$ for the same temperature function $T_\tau (\frac{1}{2})$ is plotted for comparison. The difference between the two functions appears clearly except for $\tau = 1$. (b) Same plots but for different positions $x$ and for a fixed ratio $\tau \equiv T_{\mathcal{R}} / T_{\mathcal{L}} = 5$. }
\end{figure}
\end{center}
Expression (\ref{sed}) manifestly differs from the Planck spectral energy density, a direct consequence of  the non equilibrium nature of the radiation at the macroscopic scale. This difference is illustrated in Fig. \ref{fig: 2graphs(a)} for different values of the ratio $\tau \equiv T_{\mathcal{R}} / T_{\mathcal{L}}$ and at a fixed position along the transmission line. The same observation   holds for a fixed value $\tau \neq 1$ while varying the position $x$ along the line (Fig.\ref{fig: 2graphs(b)}). 

\begin{center}
\begin{figure}[h]
\begin{centering}
\includegraphics[viewport=10bp 27bp 835bp 540bp,clip,width=1\columnwidth]{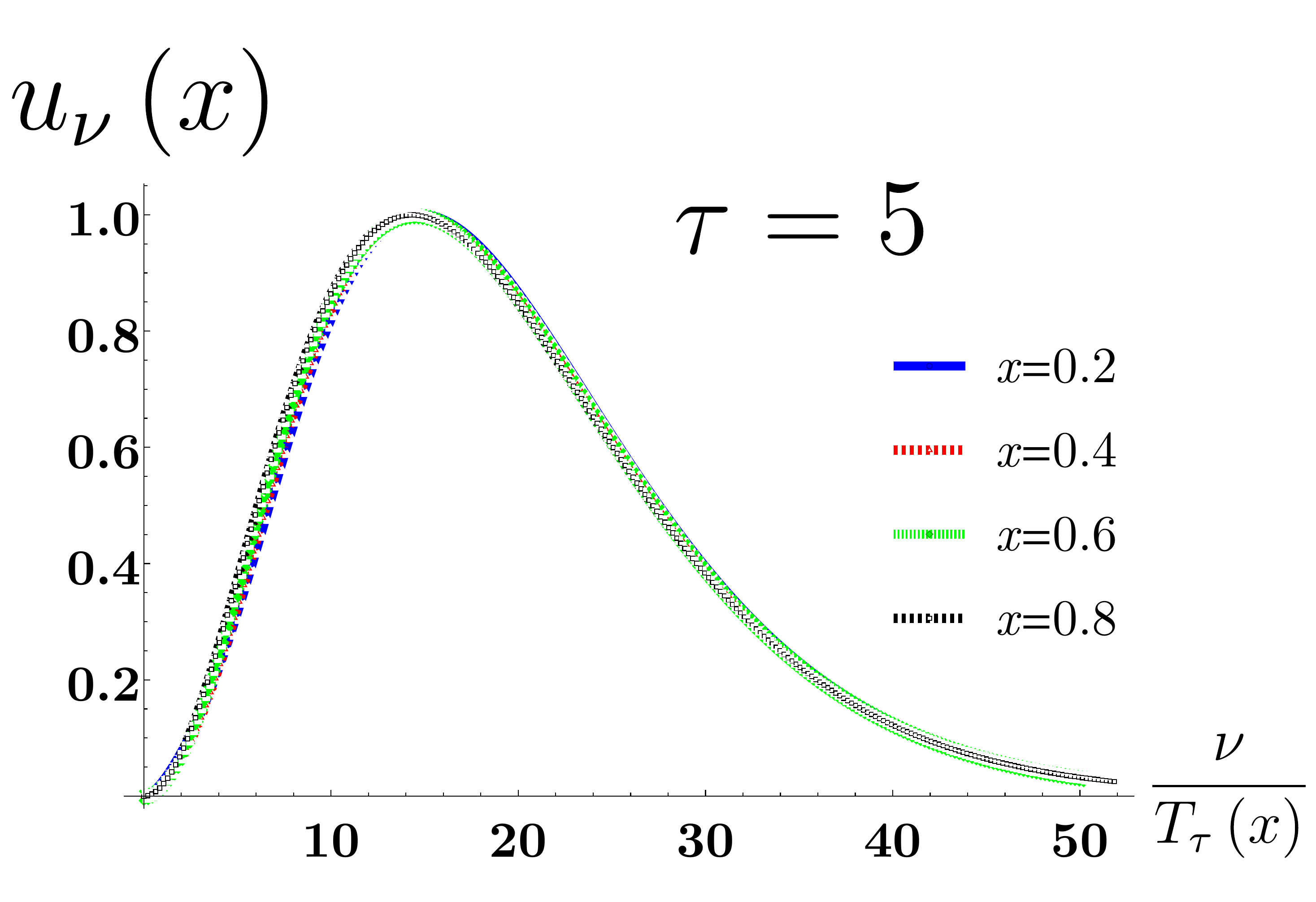}
\end{centering}
\caption{\label{fig: scaling} Scaling form (\ref{eq: scalingform}) for  $u_{ \nu}\left( x \right)$. All curves evaluated at a fixed temperature  ratio $\tau =\frac{T_{\mathcal{R}}}{T_{\mathcal{L}}}=5$ but for different values $x$ of the position, collapse to a single curve upon a proper rescaling $\nu/T_\tau \left(x\right)$ thus defining the temperature function $T_\tau (x)$.}
\end{figure}
\end{center}

A remarkable scaling form,
\begin{equation}
    u_{\nu}\left(x \right) \equiv \frac{8\pi h}{c^3} \nu^3 \Phi \left(  \nu /  T_\tau (x) \right)
    \label{eq: scalingform}
\end{equation}
for $u_{\nu}\left(x\right)$ is  observed in Fig.\ref{fig: scaling}  where the function $T_\tau (x)$, a temperature, is to be determined. It is interesting to note that while $u_{\nu}\left(x \right)$ is not a Planck distribution for $\tau \neq 1$, the scaling form (\ref{eq: scalingform}) implies  $\int \mathrm{d} \nu \, u_{\nu}\left(x \right) \propto T_\tau ^4(x) $, a behaviour reminiscent of the thermodynamic result. To understand these results and to determine the temperature $T_\tau (x)$, we now propose a fluctuating hydrodynamic description.

In the limit $L \rightarrow \infty$, upon rescaling space,  $x\rightarrow x/L$ and time, $t\rightarrow t/L^{2}$, the evolution of the stochastic model (\ref{mu},\ref{eq: product}) can be described using a fluctuating hydrodynamic Langevin equation (\ref{eq:Langevin heat current}) relating a current density $j (x, t)$ to the fluctuating local spectral energy density $u_{\nu}\left(x,t\right)$, both being constrained by the continuity equation (\ref{continuity}). 
The validity of this fluctuating hydrodynamic description, a.k.a macroscopic fluctuation theory (MFT) \citep{Bertini2015}, relies on the assumption of local equilibrium around each cell $\ell$ at an intermediate hydrodynamic scale  $(\ell \ll \ell_h \ll L)$ and for times much larger than $\ell_h ^2$ and much smaller than $L^2$, where the spectral energy density is $u_{\nu}\left(x \right)$ given in (\ref{sed}). This assumption implies that only linear response coefficients $D\left(\overline{u}_{\nu}\right)$ and $\sigma\left(\overline{u}_{\nu} \right)$ show up in the Langevin equation (\ref{eq:Langevin heat current}). To calculate them, we use the cumulant generation function $\mu^{\parallel} (\lambda)$ in (\ref{muperp}) of the total radiation current 
\begin{equation}
    Q_t = \frac{L^2}{t} \int_0^t d t' \int_0 ^1 dx \, j (x,t')
    \label{appeq:tot_int_current} 
\end{equation}
transferred between the reservoirs in a time window $\left[0,t\right]$. 
To calculate the transport coefficients $\sigma\left(\overline{u}_{\nu}\right)$ 
and $D\left(\overline{u}_{\nu}\right)$ in (\ref{eq:transp_coeff_via_SED}), we note that the Gallavotti and Cohen relation (\ref{GC}) generalises local detailed balance conditions (\ref{detailedbalance}) and allows to recover the fluctuation-dissipation theorem in the limit $E \rightarrow 0$. Hence, expanding the generating function $\mu^{\parallel} (\lambda)$ close to equilibrium, i.e. for $\tau \simeq 1$ and by setting $u_{\mathcal{L}}\equiv u,\, u_{\mathcal{R}}\equiv u +  \Delta u $ with $\Delta u \ll 1$, leads to,
\begin{subequations}\label{eq: transport coefficients}
  \begin{alignat}{2}
    \frac{\left\langle Q (t) \right\rangle }{t} = & \frac{\mathrm{d} \mu^{\parallel} (\lambda) }{\mathrm{d} \lambda} |_{\lambda = 0 } \equiv D\left(u\right)\frac{\Delta u}{L} \label{eq: diffusion}\\ 
    \frac{\left\langle Q^{2} (t)\right\rangle }{t} = & \frac{\mathrm{d}^2 \mu^{\parallel} (\lambda) }{\mathrm{d} \lambda^2} |_{\lambda = 0 } \equiv \frac{\sigma\left(u\right)}{L}\;\label{eq: sigma conductivity}.
\end{alignat}
\end{subequations}
The resulting expressions (\ref{eq:transp_coeff_via_SED}) for $D\left(\overline{u}_{\nu}\left(x\right)\right)$ and $\sigma\left(\overline{u}_{\nu} \left(x\right)\right)$ coincide with those established for the ZRP \citep{de1984remark,Landim1997,Schutz2007,Hirschberg2015} (S\,V of \citep{suppmat1}). An elementary consequence \citep{Derrida2007} of the generalised detailed balance relation (\ref{GC}),
implies that they abide the Einstein relation, 
\begin{equation}
    \frac{2D\left(\overline{u}_{\nu}\right)}{\sigma\left(\overline{u}_{\nu}\right)}= - \, \overline{s}_{\nu}^{\prime\prime} , \label{eq: einstein relation}
\end{equation}
relating $D$ and $\sigma$ to the second derivative of the local spectral entropy density $\overline{s}_{\nu}\left(\overline{u}_{\nu} (x)\right)\equiv s_\nu / \frac{8 \pi k_B\nu^2}{c^3}$ with respect to $\overline{u}_{\nu}$.

\begin{center}
\begin{figure}[h]
\begin{centering}
\includegraphics[viewport=30bp 0bp 413bp 270bp,clip,width=1\columnwidth]{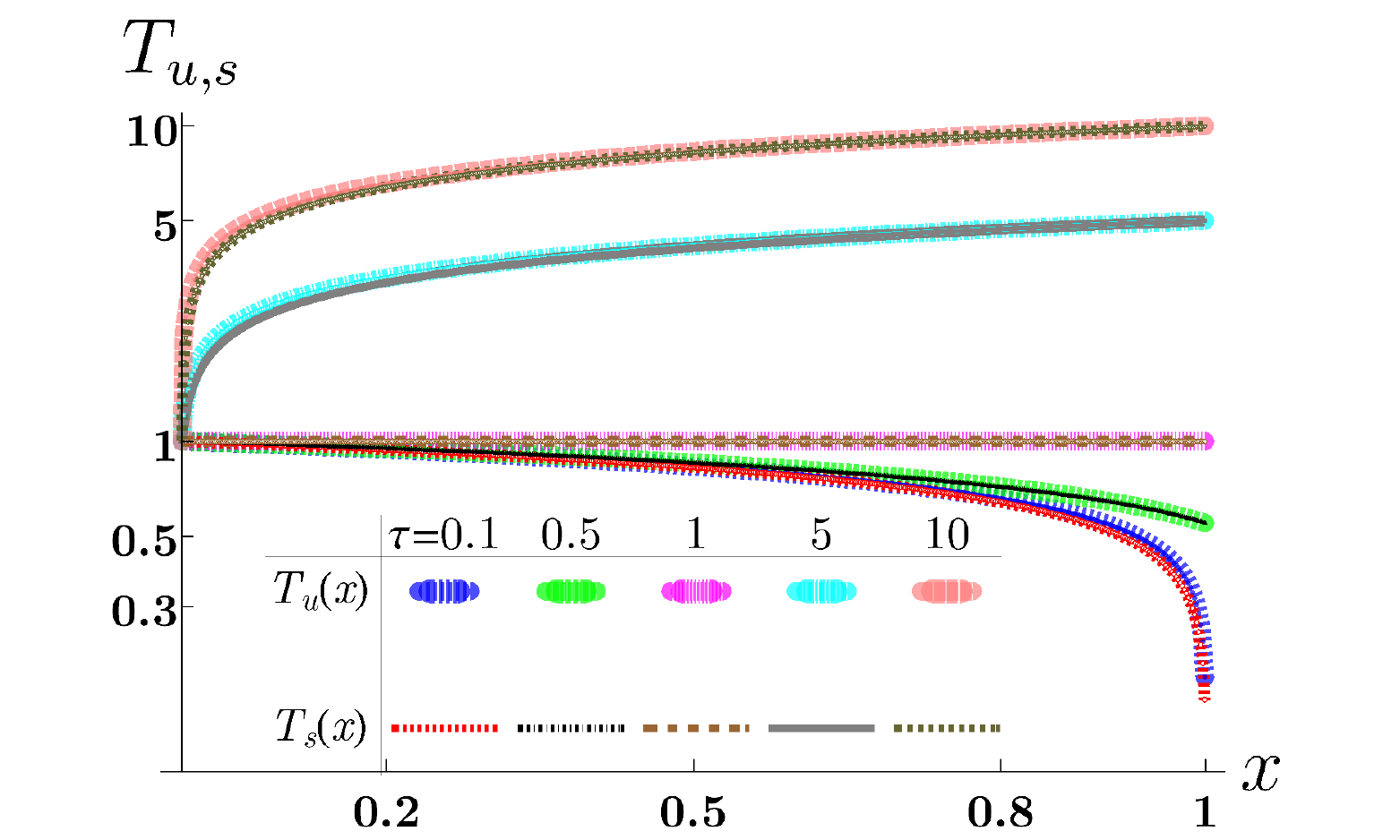}
\end{centering}
\caption{\label{fig:3d temperature profile} Comparison between the temperatures $T_u$ and $T_s$ defined from the energy and entropy densities $u(x)$ and $s(x)$, as functions of the location $x$ along the transmission line.}
\end{figure}
\end{center}

Einstein relation (\ref{eq: einstein relation}) is useful since it allows to calculate $s_\nu (x)$ by a direct integration and using (\ref{eq: scalingform}). This leads to
\begin{equation}
    s_\nu (x) = - \frac{8 \pi k_B}{c^3}\nu^2 \left[ \left( 1 + \Phi  \right) \ln \left( 1 + \Phi  \right) - \Phi  \ln \Phi   \right] \, .
\label{snu} 
\end{equation}
Since $\Phi $ depends only on the argument $ \nu /  T_\tau (x) $, then $s ( x) = \int_0 ^\infty \mathrm{d} \nu \, s_\nu (x) =  A_{s} T_{s} ^3 (x)$, where $A_{s}$ is a constant which depends on $\Phi$. The scaling form (\ref{eq: scalingform}) implies $u ( x) = \int_0 ^\infty \mathrm{d} \nu \, u_\nu (x) = A_{u} \, T_{u} ^4 (x)$, where $A_u$  is another constant which depends on $\Phi (x)$. From these two relations, we recover the familiar thermodynamic relation  $\partial s(x) / \partial u(x) \propto  1 / T_\tau (x)$, so that  $T_\tau (x)$ is indeed a local temperature \footnote{The proportionality constant is $A_{T}$, i.e $\partial s(x) / \partial u(x) =  A_{T} / T_\tau (x)$. See S\,VI of \citep{suppmat1} for the proof of the relation $A_T=\frac{3A_{s}}{4A_{u}}$. In contrast, at equilibrium, $\tau=1$, the constants are $\frac{A_{s,\text{eq}}}{A_{u,\text{eq}}}=\frac{4}{3}$, such that  $A_{T,\text{eq}}=1$ }. This result has been checked numerically in Fig.\ref{fig:3d temperature profile} where the two temperatures $T_u$ and $T_s$, respectively retrieved from $u_{\nu}(x)$ in (\ref{eq: scalingform}) and from $s_{\nu}(x)$ in (\ref{snu}), are shown to coincide and to be equal to $T_\tau (x)$. 

To summarize our findings, we have proposed a macroscopic hydrodynamic description of non equilibrium radiation. We have considered the workable example where the radiation is driven out of equilibrium by thermal coupling to two blackbody reservoirs at different temperatures. Yet, the generality of our findings appears to be independent of this specific model. We have shown that our boundary driven lattice gas model based on the bookkeeping of local photon exchanges, shares the universality of the ZRP model \citep{Levine2005,de1984remark,Schutz2007,Hirschberg2015}. Moreover, a useful macroscopic hydrodynamic limit described by the Langevin equation (\ref{eq:Langevin heat current}) has been obtained which depends on two transport parameters $D(u_{\nu})$ and $\sigma (u_{\nu})$ only. These results constitute an additional contribution to an already abundant literature on stationary, out of equilibrium, boundary driven systems \citep{Landim1997,Bertini2006,Harris2005,Hirschberg2015}. In the present case, the starting point of our study is a rather general scheme for non equilibrium photon propagation which could be further extended to other sources of fluctuating light either coherent or incoherent (e.g. lasers). Our results constitute a starting point to study more complicated situations such as the action of non equilibrium radiation on atomic motion (e.g. optical tweezers or atomic cooling), or the dynamics of quantum entanglement between two quantum particles interacting with non equilibrium light. Dynamical phase transitions and the existence of some form of condensation in the ZRP are interesting directions worth pursuing. Another interesting question is the control of fluctuating quantities in the hydrodynamic description, e.g. by means of "Thermodynamic Uncertainty Relation" recently proposed \citep{Barato2015}.

\section*{Acknowledgments}
\begin{acknowledgments}
This work was supported by the Israel Science Foundation Grant No.~772/21 and by the Pazy Foundation.
\end{acknowledgments}.


\begin{thebibliography}{32}%
\makeatletter
\providecommand \@ifxundefined [1]{%
 \@ifx{#1\undefined}
}%
\providecommand \@ifnum [1]{%
 \ifnum #1\expandafter \@firstoftwo
 \else \expandafter \@secondoftwo
 \fi
}%
\providecommand \@ifx [1]{%
 \ifx #1\expandafter \@firstoftwo
 \else \expandafter \@secondoftwo
 \fi
}%
\providecommand \natexlab [1]{#1}%
\providecommand \enquote  [1]{``#1''}%
\providecommand \bibnamefont  [1]{#1}%
\providecommand \bibfnamefont [1]{#1}%
\providecommand \citenamefont [1]{#1}%
\providecommand \href@noop [0]{\@secondoftwo}%
\providecommand \href [0]{\begingroup \@sanitize@url \@href}%
\providecommand \@href[1]{\@@startlink{#1}\@@href}%
\providecommand \@@href[1]{\endgroup#1\@@endlink}%
\providecommand \@sanitize@url [0]{\catcode `\\12\catcode `\$12\catcode
  `\&12\catcode `\#12\catcode `\^12\catcode `\_12\catcode `\%12\relax}%
\providecommand \@@startlink[1]{}%
\providecommand \@@endlink[0]{}%
\providecommand \url  [0]{\begingroup\@sanitize@url \@url }%
\providecommand \@url [1]{\endgroup\@href {#1}{\urlprefix }}%
\providecommand \urlprefix  [0]{URL }%
\providecommand \Eprint [0]{\href }%
\providecommand \doibase [0]{http://dx.doi.org/}%
\providecommand \selectlanguage [0]{\@gobble}%
\providecommand \bibinfo  [0]{\@secondoftwo}%
\providecommand \bibfield  [0]{\@secondoftwo}%
\providecommand \translation [1]{[#1]}%
\providecommand \BibitemOpen [0]{}%
\providecommand \bibitemStop [0]{}%
\providecommand \bibitemNoStop [0]{.\EOS\space}%
\providecommand \EOS [0]{\spacefactor3000\relax}%
\providecommand \BibitemShut  [1]{\csname bibitem#1\endcsname}%
\let\auto@bib@innerbib\@empty
\bibitem [{\citenamefont {Planck}(1900)}]{planck1900theory}%
  \BibitemOpen
  \bibfield  {author} {\bibinfo {author} {\bibfnamefont {M.}~\bibnamefont
  {Planck}},\ }\href@noop {} {\bibfield  {journal} {\bibinfo  {journal} {Verh.
  Deut. Phys. Ges}\ }\textbf {\bibinfo {volume} {2}},\ \bibinfo {pages} {237}
  (\bibinfo {year} {1900})}\BibitemShut {NoStop}%
\bibitem [{\citenamefont {Einstein}(1917)}]{Einstein1917}%
  \BibitemOpen
  \bibfield  {author} {\bibinfo {author} {\bibfnamefont {A.}~\bibnamefont
  {Einstein}},\ }\href@noop {} {\bibfield  {journal} {\bibinfo  {journal}
  {Verh. Deut. Phys. Ges.}\ }\textbf {\bibinfo {volume} {13}},\ \bibinfo
  {pages} {318} (\bibinfo {year} {1917})}\BibitemShut {NoStop}%
\bibitem [{\citenamefont {Glauber}(1963)}]{glauber1963coherent}%
  \BibitemOpen
  \bibfield  {author} {\bibinfo {author} {\bibfnamefont {R.~J.}\ \bibnamefont
  {Glauber}},\ }\href {\doibase 10.1103/PhysRev.131.2766} {\bibfield  {journal}
  {\bibinfo  {journal} {Phys. Rev.}\ }\textbf {\bibinfo {volume} {131}},\
  \bibinfo {pages} {2766} (\bibinfo {year} {1963})}\BibitemShut {NoStop}%
\bibitem [{\citenamefont {Titulaer}\ and\ \citenamefont
  {Glauber}(1966)}]{titulaer1966density}%
  \BibitemOpen
  \bibfield  {author} {\bibinfo {author} {\bibfnamefont {U.~M.}\ \bibnamefont
  {Titulaer}}\ and\ \bibinfo {author} {\bibfnamefont {R.~J.}\ \bibnamefont
  {Glauber}},\ }\href {\doibase 10.1103/PhysRev.145.1041} {\bibfield  {journal}
  {\bibinfo  {journal} {Phys. Rev.}\ }\textbf {\bibinfo {volume} {145}},\
  \bibinfo {pages} {1041} (\bibinfo {year} {1966})}\BibitemShut {NoStop}%
\bibitem [{\citenamefont {Bra\'{n}czyk}, \citenamefont {Chenu},\ and\
  \citenamefont {Sipe}(2017)}]{branczyk2017thermal}%
  \BibitemOpen
  \bibfield  {author} {\bibinfo {author} {\bibfnamefont {A.~M.}\ \bibnamefont
  {Bra\'{n}czyk}}, \bibinfo {author} {\bibfnamefont {A.}~\bibnamefont {Chenu}},
  \ and\ \bibinfo {author} {\bibfnamefont {J.~E.}\ \bibnamefont {Sipe}},\
  }\href {\doibase 10.1364/JOSAB.34.001536} {\bibfield  {journal} {\bibinfo
  {journal} {J. Opt. Soc. Am. B}\ }\textbf {\bibinfo {volume} {34}},\ \bibinfo
  {pages} {1536} (\bibinfo {year} {2017})}\BibitemShut {NoStop}%
\bibitem [{\citenamefont {Kindermann}, \citenamefont {Nazarov},\ and\
  \citenamefont {Beenakker}(2002)}]{kindermann2002manipulation}%
  \BibitemOpen
  \bibfield  {author} {\bibinfo {author} {\bibfnamefont {M.}~\bibnamefont
  {Kindermann}}, \bibinfo {author} {\bibfnamefont {Y.~V.}\ \bibnamefont
  {Nazarov}}, \ and\ \bibinfo {author} {\bibfnamefont {C.~W.~J.}\ \bibnamefont
  {Beenakker}},\ }\href {\doibase 10.1103/PhysRevLett.88.063601} {\bibfield
  {journal} {\bibinfo  {journal} {Phys. Rev. Lett.}\ }\textbf {\bibinfo
  {volume} {88}},\ \bibinfo {pages} {063601} (\bibinfo {year}
  {2002})}\BibitemShut {NoStop}%
\bibitem [{\citenamefont {Chen}(2000)}]{Chen2000}%
  \BibitemOpen
  \bibfield  {author} {\bibinfo {author} {\bibfnamefont {G.}~\bibnamefont
  {Chen}},\ }\href {\doibase 10.1023/A:1010003718481} {\bibfield  {journal}
  {\bibinfo  {journal} {Journal of Nanoparticle Research}\ }\textbf {\bibinfo
  {volume} {2}},\ \bibinfo {pages} {199} (\bibinfo {year} {2000})}\BibitemShut
  {NoStop}%
\bibitem [{\citenamefont {Joulain}\ \emph {et~al.}(2005)\citenamefont
  {Joulain}, \citenamefont {Mulet}, \citenamefont {Marquier}, \citenamefont
  {Carminati},\ and\ \citenamefont {Greffet}}]{Joulain2005}%
  \BibitemOpen
  \bibfield  {author} {\bibinfo {author} {\bibfnamefont {K.}~\bibnamefont
  {Joulain}}, \bibinfo {author} {\bibfnamefont {J.~P.}\ \bibnamefont {Mulet}},
  \bibinfo {author} {\bibfnamefont {F.}~\bibnamefont {Marquier}}, \bibinfo
  {author} {\bibfnamefont {R.}~\bibnamefont {Carminati}}, \ and\ \bibinfo
  {author} {\bibfnamefont {J.~J.}\ \bibnamefont {Greffet}},\ }\href {\doibase
  10.1016/j.surfrep.2004.12.002} {\bibfield  {journal} {\bibinfo  {journal}
  {Surface Science Reports}\ }\textbf {\bibinfo {volume} {57}},\ \bibinfo
  {pages} {59} (\bibinfo {year} {2005})}\BibitemShut {NoStop}%
\bibitem [{\citenamefont {Cleuren}\ and\ \citenamefont {den
  Broeck}(2007)}]{Cleuren2007}%
  \BibitemOpen
  \bibfield  {author} {\bibinfo {author} {\bibfnamefont {B.}~\bibnamefont
  {Cleuren}}\ and\ \bibinfo {author} {\bibfnamefont {C.}~\bibnamefont {den
  Broeck}},\ }\href {\doibase 10.1209/0295-5075/79/30001} {\bibfield  {journal}
  {\bibinfo  {journal} {Epl}\ }\textbf {\bibinfo {volume} {79}},\ \bibinfo
  {pages} {30001} (\bibinfo {year} {2007})}\BibitemShut {NoStop}%
\bibitem [{\citenamefont {Biehs}, \citenamefont {Rousseau},\ and\ \citenamefont
  {Greffet}(2010)}]{Biehs2010}%
  \BibitemOpen
  \bibfield  {author} {\bibinfo {author} {\bibfnamefont {S.~A.}\ \bibnamefont
  {Biehs}}, \bibinfo {author} {\bibfnamefont {E.}~\bibnamefont {Rousseau}}, \
  and\ \bibinfo {author} {\bibfnamefont {J.~J.}\ \bibnamefont {Greffet}},\
  }\href {\doibase 10.1103/PhysRevLett.105.234301} {\bibfield  {journal}
  {\bibinfo  {journal} {Physical Review Letters}\ }\textbf {\bibinfo {volume}
  {105}},\ \bibinfo {pages} {3} (\bibinfo {year} {2010})}\BibitemShut {NoStop}%
\bibitem [{\citenamefont {Bunin}\ \emph {et~al.}(2013)\citenamefont {Bunin},
  \citenamefont {Kafri}, \citenamefont {Lecomte}, \citenamefont {Podolsky},\
  and\ \citenamefont {Polkovnikov}}]{bunin2013transport}%
  \BibitemOpen
  \bibfield  {author} {\bibinfo {author} {\bibfnamefont {G.}~\bibnamefont
  {Bunin}}, \bibinfo {author} {\bibfnamefont {Y.}~\bibnamefont {Kafri}},
  \bibinfo {author} {\bibfnamefont {V.}~\bibnamefont {Lecomte}}, \bibinfo
  {author} {\bibfnamefont {D.}~\bibnamefont {Podolsky}}, \ and\ \bibinfo
  {author} {\bibfnamefont {A.}~\bibnamefont {Polkovnikov}},\ }\href {\doibase
  10.1088/1742-5468/2013/08/P08015} {\bibfield  {journal} {\bibinfo  {journal}
  {Journal of Statistical Mechanics: Theory and Experiment}\ }\textbf {\bibinfo
  {volume} {2013}},\ \bibinfo {pages} {P08015} (\bibinfo {year}
  {2013})}\BibitemShut {NoStop}%
\bibitem [{\citenamefont {Nicacio}\ \emph {et~al.}(2015)\citenamefont
  {Nicacio}, \citenamefont {Ferraro}, \citenamefont {Imparato}, \citenamefont
  {Paternostro},\ and\ \citenamefont {Semi{\~{a}}o}}]{Nicacio2015}%
  \BibitemOpen
  \bibfield  {author} {\bibinfo {author} {\bibfnamefont {F.}~\bibnamefont
  {Nicacio}}, \bibinfo {author} {\bibfnamefont {A.}~\bibnamefont {Ferraro}},
  \bibinfo {author} {\bibfnamefont {A.}~\bibnamefont {Imparato}}, \bibinfo
  {author} {\bibfnamefont {M.}~\bibnamefont {Paternostro}}, \ and\ \bibinfo
  {author} {\bibfnamefont {F.~L.}\ \bibnamefont {Semi{\~{a}}o}},\ }\href
  {\doibase 10.1103/PhysRevE.91.042116} {\bibfield  {journal} {\bibinfo
  {journal} {Physical Review E - Statistical, Nonlinear, and Soft Matter
  Physics}\ }\textbf {\bibinfo {volume} {91}} (\bibinfo {year} {2015}),\
  10.1103/PhysRevE.91.042116}\BibitemShut {NoStop}%
\bibitem [{\citenamefont {Greffet}\ \emph {et~al.}(2018)\citenamefont
  {Greffet}, \citenamefont {Bouchon}, \citenamefont {Brucoli},\ and\
  \citenamefont {Marquier}}]{Greffet2018}%
  \BibitemOpen
  \bibfield  {author} {\bibinfo {author} {\bibfnamefont {J.~J.}\ \bibnamefont
  {Greffet}}, \bibinfo {author} {\bibfnamefont {P.}~\bibnamefont {Bouchon}},
  \bibinfo {author} {\bibfnamefont {G.}~\bibnamefont {Brucoli}}, \ and\
  \bibinfo {author} {\bibfnamefont {F.}~\bibnamefont {Marquier}},\ }\href
  {\doibase 10.1103/PhysRevX.8.021008} {\bibfield  {journal} {\bibinfo
  {journal} {Physical Review X}\ }\textbf {\bibinfo {volume} {8}},\ \bibinfo
  {pages} {21008} (\bibinfo {year} {2018})}\BibitemShut {NoStop}%
\bibitem [{\citenamefont {Lepri}, \citenamefont {Livi},\ and\ \citenamefont
  {Politi}(2003)}]{LEPRI20031}%
  \BibitemOpen
  \bibfield  {author} {\bibinfo {author} {\bibfnamefont {S.}~\bibnamefont
  {Lepri}}, \bibinfo {author} {\bibfnamefont {R.}~\bibnamefont {Livi}}, \ and\
  \bibinfo {author} {\bibfnamefont {A.}~\bibnamefont {Politi}},\ }\href
  {\doibase https://doi.org/10.1016/S0370-1573(02)00558-6} {\bibfield
  {journal} {\bibinfo  {journal} {Physics Reports}\ }\textbf {\bibinfo {volume}
  {377}},\ \bibinfo {pages} {1} (\bibinfo {year} {2003})}\BibitemShut {NoStop}%
\bibitem [{\citenamefont {Bernard}\ and\ \citenamefont
  {Doyon}(2012)}]{Bernard2012}%
  \BibitemOpen
  \bibfield  {author} {\bibinfo {author} {\bibfnamefont {D.}~\bibnamefont
  {Bernard}}\ and\ \bibinfo {author} {\bibfnamefont {B.}~\bibnamefont
  {Doyon}},\ }\href {\doibase 10.1088/1751-8113/45/36/362001} {\bibfield
  {journal} {\bibinfo  {journal} {Journal of Physics A: Mathematical and
  Theoretical}\ }\textbf {\bibinfo {volume} {45}},\ \bibinfo {pages} {362001}
  (\bibinfo {year} {2012})}\BibitemShut {NoStop}%
\bibitem [{\citenamefont {Xuereb}, \citenamefont {Imparato},\ and\
  \citenamefont {Dantan}(2015)}]{Xuereb2015}%
  \BibitemOpen
  \bibfield  {author} {\bibinfo {author} {\bibfnamefont {A.}~\bibnamefont
  {Xuereb}}, \bibinfo {author} {\bibfnamefont {A.}~\bibnamefont {Imparato}}, \
  and\ \bibinfo {author} {\bibfnamefont {A.}~\bibnamefont {Dantan}},\ }\href
  {\doibase 10.1088/1367-2630/17/5/055013} {\bibfield  {journal} {\bibinfo
  {journal} {New Journal of Physics}\ }\textbf {\bibinfo {volume} {17}}
  (\bibinfo {year} {2015}),\ 10.1088/1367-2630/17/5/055013}\BibitemShut
  {NoStop}%
\bibitem [{\citenamefont {Evans}\ and\ \citenamefont
  {Hanney}(2005)}]{Evans2005}%
  \BibitemOpen
  \bibfield  {author} {\bibinfo {author} {\bibfnamefont {M.~R.}\ \bibnamefont
  {Evans}}\ and\ \bibinfo {author} {\bibfnamefont {T.}~\bibnamefont {Hanney}},\
  }\href {\doibase 10.1088/0305-4470/38/19/R01} {\bibfield  {journal} {\bibinfo
   {journal} {J. Phys. A: Math. Gen.}\ }\textbf {\bibinfo {volume} {38}}
  (\bibinfo {year} {2005}),\ 10.1088/0305-4470/38/19/R01}\BibitemShut {NoStop}%
\bibitem [{\citenamefont {Bertini}\ \emph {et~al.}(2015)\citenamefont
  {Bertini}, \citenamefont {{De Sole}}, \citenamefont {Gabrielli},
  \citenamefont {Jona-Lasinio},\ and\ \citenamefont {Landim}}]{Bertini2015}%
  \BibitemOpen
  \bibfield  {author} {\bibinfo {author} {\bibfnamefont {L.}~\bibnamefont
  {Bertini}}, \bibinfo {author} {\bibfnamefont {A.}~\bibnamefont {{De Sole}}},
  \bibinfo {author} {\bibfnamefont {D.}~\bibnamefont {Gabrielli}}, \bibinfo
  {author} {\bibfnamefont {G.}~\bibnamefont {Jona-Lasinio}}, \ and\ \bibinfo
  {author} {\bibfnamefont {C.}~\bibnamefont {Landim}},\ }\href {\doibase
  10.1103/RevModPhys.87.593} {\bibfield  {journal} {\bibinfo  {journal}
  {Reviews of Modern Physics}\ }\textbf {\bibinfo {volume} {87}},\ \bibinfo
  {pages} {593} (\bibinfo {year} {2015})}\BibitemShut {NoStop}%
\bibitem [{\citenamefont {Derrida}(2007)}]{Derrida2007}%
  \BibitemOpen
  \bibfield  {author} {\bibinfo {author} {\bibfnamefont {B.}~\bibnamefont
  {Derrida}},\ }\href {\doibase 10.1088/1742-5468/2007/07/P07023} {\bibfield
  {journal} {\bibinfo  {journal} {J. Stat. Mech.}\ }\textbf {\bibinfo {volume}
  {07023}},\ \bibinfo {pages} {P07023} (\bibinfo {year} {2007})}\BibitemShut
  {NoStop}%
\bibitem [{sup()}]{suppmat1}%
  \BibitemOpen
  \href@noop {} {}\bibinfo {howpublished} {Suplementary material is available
  at \url{URL_will_be_inserted_by_publisher}}\BibitemShut {NoStop}%
\bibitem [{\citenamefont {Harris}, \citenamefont {R{\'{a}}kos},\ and\
  \citenamefont {Sch{\"{u}}tz}(2005)}]{Harris2005}%
  \BibitemOpen
  \bibfield  {author} {\bibinfo {author} {\bibfnamefont {R.~J.}\ \bibnamefont
  {Harris}}, \bibinfo {author} {\bibfnamefont {A.}~\bibnamefont {R{\'{a}}kos}},
  \ and\ \bibinfo {author} {\bibfnamefont {G.~M.}\ \bibnamefont
  {Sch{\"{u}}tz}},\ }\href {\doibase 10.1088/1742-5468/2005/08/P08003}
  {\bibfield  {journal} {\bibinfo  {journal} {Journal of Statistical Mechanics:
  Theory and Experiment}\ ,\ \bibinfo {pages} {55}} (\bibinfo {year}
  {2005})}\BibitemShut {NoStop}%
\bibitem [{\citenamefont {Levine}, \citenamefont {Mukamel},\ and\ \citenamefont
  {Sch{\"{u}}tz}(2005)}]{Levine2005}%
  \BibitemOpen
  \bibfield  {author} {\bibinfo {author} {\bibfnamefont {E.}~\bibnamefont
  {Levine}}, \bibinfo {author} {\bibfnamefont {D.}~\bibnamefont {Mukamel}}, \
  and\ \bibinfo {author} {\bibfnamefont {G.~M.}\ \bibnamefont {Sch{\"{u}}tz}},\
  }\href {\doibase 10.1007/s10955-005-7000-7} {\bibfield  {journal} {\bibinfo
  {journal} {J. Stat. Phys.}\ }\textbf {\bibinfo {volume} {120}},\ \bibinfo
  {pages} {759} (\bibinfo {year} {2005})}\BibitemShut {NoStop}%
\bibitem [{Note1()}]{Note1}%
  \BibitemOpen
  \bibinfo {note} {Having $0<z_{0/L+1}<1$ imposes $\gamma >\alpha $ and $\beta
  >\delta $.}\BibitemShut {Stop}%
\bibitem [{\citenamefont {Gallavotti}\ and\ \citenamefont
  {Cohen}(1995)}]{Gallavotti1995}%
  \BibitemOpen
  \bibfield  {author} {\bibinfo {author} {\bibfnamefont {G.}~\bibnamefont
  {Gallavotti}}\ and\ \bibinfo {author} {\bibfnamefont {E.~G.~D.}\ \bibnamefont
  {Cohen}},\ }\href {\doibase 10.1103/PhysRevLett.74.2694} {\bibfield
  {journal} {\bibinfo  {journal} {Physical Review Letters}\ }\textbf {\bibinfo
  {volume} {74}},\ \bibinfo {pages} {2694} (\bibinfo {year}
  {1995})}\BibitemShut {NoStop}%
\bibitem [{\citenamefont {Lebowitz}\ and\ \citenamefont
  {Spohn}(1998)}]{Lebowitz1998}%
  \BibitemOpen
  \bibfield  {author} {\bibinfo {author} {\bibfnamefont {J.~L.}\ \bibnamefont
  {Lebowitz}}\ and\ \bibinfo {author} {\bibfnamefont {H.}~\bibnamefont
  {Spohn}},\ }\href {\doibase 10.1023/A:1004589714161} {\bibfield  {journal}
  {\bibinfo  {journal} {Journal of Statistical Physics}\ ,\ \bibinfo {pages}
  {333}} (\bibinfo {year} {1998})}\BibitemShut {NoStop}%
\bibitem [{\citenamefont {De~Masi}\ and\ \citenamefont
  {Ferrari}(1984)}]{de1984remark}%
  \BibitemOpen
  \bibfield  {author} {\bibinfo {author} {\bibfnamefont {A.}~\bibnamefont
  {De~Masi}}\ and\ \bibinfo {author} {\bibfnamefont {P.}~\bibnamefont
  {Ferrari}},\ }\href {https://link.springer.com/article/10.1007/BF01015727}
  {\bibfield  {journal} {\bibinfo  {journal} {Journal of statistical physics}\
  }\textbf {\bibinfo {volume} {36}},\ \bibinfo {pages} {81} (\bibinfo {year}
  {1984})}\BibitemShut {NoStop}%
\bibitem [{\citenamefont {Landim}\ and\ \citenamefont
  {Mourragui}(1997)}]{Landim1997}%
  \BibitemOpen
  \bibfield  {author} {\bibinfo {author} {\bibfnamefont {C.}~\bibnamefont
  {Landim}}\ and\ \bibinfo {author} {\bibfnamefont {M.}~\bibnamefont
  {Mourragui}},\ }\href {\doibase
  https://doi.org/10.1016/S0246-0203(97)80116-1} {\bibfield  {journal}
  {\bibinfo  {journal} {Annales de l'institut Henri Poincare (B) Probability
  and Statistics}\ }\textbf {\bibinfo {volume} {33}},\ \bibinfo {pages} {65}
  (\bibinfo {year} {1997})}\BibitemShut {NoStop}%
\bibitem [{\citenamefont {Sch{\"{u}}tz}\ and\ \citenamefont
  {Harris}(2007)}]{Schutz2007}%
  \BibitemOpen
  \bibfield  {author} {\bibinfo {author} {\bibfnamefont {G.~M.}\ \bibnamefont
  {Sch{\"{u}}tz}}\ and\ \bibinfo {author} {\bibfnamefont {R.~J.}\ \bibnamefont
  {Harris}},\ }\href {\doibase 10.1007/s10955-007-9280-6} {\bibfield  {journal}
  {\bibinfo  {journal} {Journal of Statistical Physics}\ }\textbf {\bibinfo
  {volume} {127}},\ \bibinfo {pages} {419} (\bibinfo {year}
  {2007})}\BibitemShut {NoStop}%
\bibitem [{\citenamefont {Hirschberg}, \citenamefont {Mukamel},\ and\
  \citenamefont {Schütz}(2015)}]{Hirschberg2015}%
  \BibitemOpen
  \bibfield  {author} {\bibinfo {author} {\bibfnamefont {O.}~\bibnamefont
  {Hirschberg}}, \bibinfo {author} {\bibfnamefont {D.}~\bibnamefont {Mukamel}},
  \ and\ \bibinfo {author} {\bibfnamefont {G.~M.}\ \bibnamefont {Schütz}},\
  }\href {\doibase 10.1088/1742-5468/2015/11/p11023} {\bibfield  {journal}
  {\bibinfo  {journal} {Journal of Statistical Mechanics: Theory and
  Experiment}\ }\textbf {\bibinfo {volume} {2015}},\ \bibinfo {pages} {P11023}
  (\bibinfo {year} {2015})}\BibitemShut {NoStop}%
\bibitem [{Note2()}]{Note2}%
  \BibitemOpen
  \bibinfo {note} {The proportionality constant is $A_{T}$, i.e $\partial s(x)
  / \partial u(x) = A_{T} / T_\tau (x)$. See S\protect \tmspace +\thinmuskip
  {.1667em}VI of \protect \citep {suppmat1} for the proof of the relation
  $A_T=\protect \frac {3A_{s}}{4A_{u}}$. In contrast, at equilibrium, $\tau
  =1$, the constants are $\protect \frac {A_{s,\protect \text
  {eq}}}{A_{u,\protect \text {eq}}}=\protect \frac {4}{3}$, such that
  $A_{T,\protect \text {eq}}=1$}\BibitemShut {NoStop}%
\bibitem [{\citenamefont {Bertini}\ \emph {et~al.}(2006)\citenamefont
  {Bertini}, \citenamefont {{De Sole}}, \citenamefont {Gabrielli},
  \citenamefont {Jona-Lasinio},\ and\ \citenamefont {Landim}}]{Bertini2006}%
  \BibitemOpen
  \bibfield  {author} {\bibinfo {author} {\bibfnamefont {L.}~\bibnamefont
  {Bertini}}, \bibinfo {author} {\bibfnamefont {A.}~\bibnamefont {{De Sole}}},
  \bibinfo {author} {\bibfnamefont {D.}~\bibnamefont {Gabrielli}}, \bibinfo
  {author} {\bibfnamefont {G.}~\bibnamefont {Jona-Lasinio}}, \ and\ \bibinfo
  {author} {\bibfnamefont {C.}~\bibnamefont {Landim}},\ }\href {\doibase
  10.1007/s10955-006-9056-4} {\bibfield  {journal} {\bibinfo  {journal}
  {Journal of Statistical Physics}\ }\textbf {\bibinfo {volume} {123}},\
  \bibinfo {pages} {237} (\bibinfo {year} {2006})}\BibitemShut {NoStop}%
\bibitem [{\citenamefont {Barato}\ and\ \citenamefont
  {Seifert}(2015)}]{Barato2015}%
  \BibitemOpen
  \bibfield  {author} {\bibinfo {author} {\bibfnamefont {A.~C.}\ \bibnamefont
  {Barato}}\ and\ \bibinfo {author} {\bibfnamefont {U.}~\bibnamefont
  {Seifert}},\ }\href {\doibase 10.1103/PhysRevLett.114.158101} {\bibfield
  {journal} {\bibinfo  {journal} {Phys. Rev. Lett.}\ }\textbf {\bibinfo
  {volume} {114}},\ \bibinfo {pages} {158101} (\bibinfo {year}
  {2015})}\BibitemShut {NoStop}%
\end{thebibliography}
%

\ifarXiv
    

\section*{Supplementary material}

# Supplemental Material: Hydrodynamic description of Non-Equilibrium Radiation

## S I. PHOTON ESCAPE RATE

In this section we present a derivation, using kinetic theory, of the relations $\alpha = 2A\,c\,u_\nu(T_0)/h$ and $\delta = 2A\,c\,u_\nu(T_{L+1})/h$, between the rates of incoming photons from the blackbody reservoirs at the boundaries, as defined and depicted in Fig.2 in the main text.

Consider a single blackbody cavity at temperature $T$, with an aperture of area $A$ from which photons escape the cavity. The escape rate is computed using the following kinematic argument. The number of photons that leave the cavity per unit time $\mathrm{d}t$ through an aperture of area $A$, are those which reside in the volume $Ac\cos\theta\,\mathrm{d}t$, i.e

$$N = A\rho_{\mathrm{ph}} c \cos\theta\,\mathrm{d}t\;,$$

where $\theta$ is the angle between the photon wave vector and the normal to the aperture. Hence, their escape rate is

$$\alpha = A \int_{-\pi/2}^{\pi/2} \mathrm{d}\theta\, \rho_{\mathrm{ph}} c \cos\theta\;.$$

The photon density depends on the density of states, the photon frequency $\nu$ and the average occupation of this frequency mode, namely

$$\rho_{\mathrm{ph}} = g_\nu \nu \langle n_\nu \rangle\;.$$

Hence the rate is

$$\alpha = 2A g_\nu \nu c \langle n_\nu \rangle\;.$$

Rewriting the above in terms of the spectral energy density $u_\nu(T) = g_\nu h \nu \langle n_\nu \rangle$, with the temperature in the reservoirs, leads to the relations given in the text.

## S II. CUMULANT GENERATING FUNCTION OF THE TOTAL RADIATION CURRENT

In this section, we present a calculation of the cumulant generating function $\mu^{\parallel}(\lambda)$ from equation (9) in the main text.

We define the total number of photons that leave the transmission line from its left edge, in a time windows $[0,t]$

$$Q^{\mathcal{L}}_{[0,t]} = \overrightarrow{1}_{[0,t]} - \overleftarrow{1}_{[0,t]}\;, \tag{S1}$$

where $\overleftarrow{1}_{[0,t]}$ is the number of photons hopping from cell 1 to the left reservoir and $\overrightarrow{1}_{[0,t]}$ is the number of photon hopping from the left reservoir to cell 1, all in the time window $[0,t]$. The quantity of interest is the cumulant generating function,

$$\mu_{\mathcal{L}}(\lambda) = \lim_{t\to\infty} \frac{1}{t} \ln \langle e^{\lambda Q^{\mathcal{L}}_{[0,t]}} \rangle\;, \tag{S2}$$

The total current of photons is a sum of elementary contributions of each photon that entered the chain of resonators between times 0 and $t$,

$$Q^{\mathcal{L}}_{[0,t]} = \sum_{k=0}^{N_t} Q^{k}_{[t_k,t]}\;, \tag{S3}$$

where $N_t$ is a random variable and $t_k$ is the time at which the $k$th photon entered the chain. Hence (see also appendices of [S1, S2]),

$$\left\langle e^{\lambda Q^{\mathcal{L}}_{[0,t]}} \right\rangle = \left\langle e^{\lambda \sum_{k=0}^{N_t} Q^{k}_{[t_k,t]}} \right\rangle = \sum_{n=0}^{\infty} \frac{1}{n!} \int \mathrm{d}t_1 \cdots \int \mathrm{d}t_n\, \alpha^n\, e^{-\alpha t} \prod_{k=0}^{n} \left\langle e^{\lambda Q^{k}_{[t_k,t]}} \right\rangle\;, \tag{S4}$$

with ordered injection times $0 \leq t_1 \leq t_2 \leq \cdots \leq t_n \leq t$ describing the Poisson process. To factorise the expectation value, we have used the assumption of independent photons, so that the calculation reduces to the single particle expectation value, $\left\langle e^{\lambda Q^k_{[t_k,t]}} \right\rangle$.

The processes $Q^k_{[t_k,t]}$ can be regarded as indicators, i.e

$$Q^k_{[t_k,t]} = \begin{cases} 0 & \text{if a photon } k \text{ has left the resonating chain} \\ 1 & \text{otherwise} \end{cases}$$

Then, we denote $\Pi_{\mathcal{L}}(t-t_k)$ the probability for a photon $k$ to leave the transmission line from the left edge. Hence,

$$\left\langle e^{\lambda Q^k_{[t_k,t]}} \right\rangle = \Pi_{\mathcal{L}}(t-t_k) + e^{\lambda}(1 - \Pi_{\mathcal{L}}(t-t_k)) \ .$$

For $t \to \infty$, with fixed $N$ and $t_k$, the photon leaves the chain almost surely, so that the probability $\Pi_{\mathcal{L}}(t-t_k)$ converges to a finite and constant value $\Pi^*_{\mathcal{L}}$. This probability is in the kernel of

$$\mathcal{P}_{\mathcal{L}} = \begin{pmatrix} 0 & \gamma & 0 & \cdots & \cdots & \cdots & \cdots & 0 \\ 0 & -\gamma - 1 & 1 & 0 & \cdots & \cdots & \cdots & 0 \\ 0 & 1 & -2 & 1 & 0 & \cdots & \cdots & 0 \\ 0 & 0 & 1 & -2 & 0 & \cdots & \cdots & 0 \\ \vdots & \vdots & 0 & 0 & \ddots & \ddots & 0 & 0 \\ \vdots & \vdots & \vdots & \vdots & \ddots & \ddots & -2 & -\beta - 1 \\ 0 & 0 & 0 & 0 & 0 & 0 & \beta & 0 \end{pmatrix}_{(N+2)\times(N+2)} .$$

Namely, defining the probability $P_k$ for a photon in cell $k$ to exit from the left, the $(N+2)$ component vector $\mathbf{P}$ is harmonic $\mathcal{P}_{\mathcal{L}}\mathbf{P} = 0$ with boundary conditions, $P_{k=0} = 1$ and $P_{k=N+1} = 0$. The solution of this linear system is,

$$\Pi^*_{\mathcal{L}} = P_1 = \frac{\gamma(\beta N + 1) - \gamma\beta}{\gamma(\beta N + 1) - \beta(\gamma - 1)} \ .$$

Thus, for $t \to \infty$, hence $t \gg t_k$ for all $k$, we obtain from (S4),

$$\left\langle e^{\lambda Q^{\mathcal{L}}_{[0,t]}} \right\rangle = \sum_{n=0}^{\infty} \frac{\alpha^n e^{-\alpha t}}{n!} t^n \left[ \Pi^*_{\mathcal{L}} + e^{\lambda}(1 - \Pi^*_{\mathcal{L}}) \right]^n = e^{\alpha t (F(\lambda) - 1)} \ ,$$

with

$$\alpha(F(\lambda) - 1) = \frac{\alpha\beta(e^{\lambda} - 1)}{\gamma(\beta N + 1) - \beta(\gamma - 1)} \equiv \mu_{\mathcal{L}}(\lambda) \ . \tag{S5}$$

Repeating an analogous calculation for the total number of photons $Q^{\mathcal{R}}_{[0,t]}$ entering from the right reservoir, leads to

$$\lim_{t\to\infty} \frac{1}{t} \ln \left\langle e^{\lambda Q^{\mathcal{R}}_{[0,t]}} \right\rangle = -\frac{\gamma\delta(1 - e^{-\lambda})}{\gamma(\beta N + 1) - \beta(\gamma - 1)} \equiv \mu_{\mathcal{R}}(\lambda) \ . \tag{S6}$$

Altogether, the total cumulant generating function $\mu^{\parallel}(\lambda) = \mu_{\mathcal{L}} - \mu_{\mathcal{R}}$ is given by (9) in the text,

$$\mu^{\parallel}(\lambda) = \frac{\alpha\beta(e^{\lambda} - 1) - \gamma\delta(1 - e^{-\lambda})}{\gamma(\beta N + 1) - \beta(\gamma - 1)} \ . \tag{S7}$$

## S III.  LOCAL EQUILIBRIUM VIA CREATION AND ANNIHILATION RATES

In this section we establish the sufficient condition (12), in the main text, for a local equilibrium, i.e

$$\frac{c_k(n_k)}{a_k(n_k+1)} = \frac{P_{\infty}(\eta_{\overline{k}})}{P_{\infty}(\eta)} = z_k \ . \tag{S8}$$

To this purpose, we consider the following rate model for equilibrium blackbody radiation.

For blackbody radiation at temperature $T$, the corresponding spectral energy density is given by the Planck distribution

$$u_\nu(T) = g_\nu h\nu \langle n_\nu(T) \rangle, \tag{S9}$$

where $g_\nu = 8\pi\nu^2/c^3$ is the density of modes of frequency $\nu$ inside the blackbody cavity, and $\langle n_\nu \rangle$ is the average occupation of these modes. It is calculated from the Bose-Einstein probability distribution $\pi^B$ (for Blackbody) to observe $n$ photons in the cavity,

$$\pi^B(n) = \left(1 - e^{-\frac{h\nu}{k_B T}}\right) e^{-\frac{h\nu}{k_B T} n}. \tag{S10}$$

Accordingly, the Boltzmann factor is

$$e^{-\frac{h\nu}{k_B T}} = \frac{\langle n_\nu \rangle}{1 + \langle n_\nu \rangle}. \tag{S11}$$

At this stage, it is interesting to introduce the fugacity $z^B$ by means of the change of variables, $\langle n_\nu \rangle = \frac{z^B}{1-z^B}$. It implies $z^B = \exp\left(-\frac{h\nu}{k_B T}\right)$ and $0 < z^B < 1$. This allows to rewrite the distribution (S10) in terms of fugacities,

$$\pi^B(n_\nu) = \left(1 - z^B\right)\left(z^B\right)^{n_\nu}. \tag{S12}$$

We now show that this distribution is the stationary probability $P_\infty^B$ of the following rate model.

Let us consider the radiation inside the cavity as a gas of photons, each carrying a momentum $h\nu/c$. The physical processes of absorption and emission of photons from the cavity walls, are described by means of the rates $a(n_\nu)$ and $c(n_\nu)$, respectively. The state of the radiation is encoded in the probability $P_t^B(n_\nu)$ to observe $n_\nu = 0, 1, 2, \ldots$ photons at time $t$ inside the cavity. This probability evolves according to the master equation,

$$\partial_t P_t^B(n_\nu) = \mathcal{L}_B\left[P_t^B(n_\nu)\right], \tag{S13}$$

where the generator $\mathcal{L}_B$ is

$$\mathcal{L}_B\left[P_t^B(n_\nu)\right] = c(n_\nu - 1) P_t^B(n_\nu - 1) + a(n_\nu + 1) P_t^B(n_\nu + 1) - P_t^B(n_\nu)\left[a(n_\nu) + c(n_\nu)\right]. \tag{S14}$$

The above equation is constrained by $a(0) = 0 = c(-1)$.

We now find the conditions on the rates $a(n_\nu)$ and $c(n_\nu)$ for which (S12) is the stationary state of (S13) [S3], i.e, for $t \to \infty$ we take

$$P_\infty^B(n_\nu) \equiv \pi^B(n_\nu). \tag{S15}$$

Stationarity is equivalent to $\mathcal{L}_B\left[P_\infty^B(n_\nu)\right] = 0$, thus we write

$$\begin{aligned}0 =& c(n_\nu - 1) P_\infty^B(n_\nu - 1) + a(n_\nu + 1) P_\infty^B(n_\nu + 1) - P_\infty^B(n_\nu)\left[a(n_\nu) + c(n_\nu)\right] \\=& P_\infty^B(n_\nu)\left(c(n_\nu - 1)\left[\frac{P_\infty^B(n_\nu - 1)}{P_\infty^B(n_\nu)} - \frac{a(n_\nu)}{c(n_\nu - 1)}\right] + a(n_\nu + 1)\left[\frac{P_\infty^B(n_\nu + 1)}{P_\infty^B(n_\nu)} - \frac{c(n_\nu)}{a(n_\nu + 1)}\right]\right).\end{aligned}$$

The right-hand-side vanishes, if and only if, (using (S15) and (S12))

$$\frac{c(n_\nu)}{a(n_\nu + 1)} = \frac{P_\infty^B(n_\nu + 1)}{P_\infty^B(n_\nu)} = z^B = e^{-\frac{h\nu}{k_B T}}. \tag{S16}$$

The generalization (S8) of (S16) is twofold: first, $z_k$ depends on the spatial index $k$ of a cell along the transmission line. Second, $z_k$ is not a Boltzmann factor in cell $k$, but it is rather given by (13) in the main text.

## S IV. MASTER EQUATION FOR THE PHOTON-GAS MODEL

The purpose of this section is to build the generator $\mathcal{L}$ used in equations (5) and (6) in the main text. Then, to prove that the stationary state is given by (10) and (11) of the main text.

## A. Construction of the Master equation generator $\mathcal{L}$

The generator $\mathcal{L}$ is a linear function of the probability $P_t(\eta)$ and it specifies the rate of flow of $P_t(\eta)$ between configurations $\eta = \{n_1, n_2, \ldots, n_N\}$. We recall that $\eta_{\bar{i},\underline{j}}$ accounts for a configuration which, relative to $\eta$, has an excess of one photon in the $i^{\text{th}}$ cell and a depletion of one photon in the $j^{\text{th}}$ cell, namely,

$$\eta_{\bar{i},\underline{j}} = \{n_1, \ldots, n_i + 1, \ldots, n_j - 1, \ldots n_N\}, \qquad n_j > 0.$$

Among the three possible processes (1,2,3) introduced in the main text, process (1) is independent of (2+3), so that the generator $\mathcal{L}$ can be written accordingly as a sum of two independent contributions $\mathcal{L} = \mathcal{L}^{\parallel} + \mathcal{L}^{\perp}$ with obvious notations.

The generator $\mathcal{L}^{\parallel}$ can be expressed as a sum of two contributions $\mathcal{L}^{\parallel} = \mathcal{L}_{\text{boundary}} + \mathcal{L}_{\text{bulk}}$. As described in the main text, $\mathcal{L}^{\parallel}$ generates the dynamics associated with the rates $\alpha, \beta, \gamma, \delta, q$ – processes 1 and 3 in the main text, and $\mathcal{L}^{\perp}$ generates the dynamics associated with the rates $a_k$ and $c_k$ – process 2 in the main text. These three generators are built from the following processes:

1. **Boundary** Transitions $\eta \underset{\gamma}{\overset{\alpha}{\rightleftharpoons}} \eta_{\bar{1}}$ correspond to probability flow from configuration $\eta$ to $\eta_{\bar{1}}$ with rate $\alpha$ and conversely with rate $\gamma$. A similar descriptions applies to transitions at $k = N$, so that the boundary generator is

$$\begin{aligned}\mathcal{L}_{\text{boundary}}[P_t(\eta)] =& \alpha\left[P_t(\eta_{\underline{1}}) - P_t(\eta)\right] + \gamma\left[P_t(\eta_{\bar{1}}) - P_t(\eta)\right] \\ &+ \delta\left[P(\eta_{\underline{N}}, t) - P_t(\eta)\right] + \beta\left[P_t(\eta_{\bar{N}}) - P_t(\eta)\right]\end{aligned} \tag{S17}$$

2. **Bulk:** Transitions $\eta_{\bar{k},\underline{k+1}} \underset{q}{\overset{q}{\rightleftharpoons}} \eta_{\underline{k},\overline{k+1}}$ reflect the assumption of symmetric photon exchange between adjacent cells in the resonating chain, namely, there is an equal probability, per unit time, for a photon to go from cell $k$ to $k+1$ and vice-versa. Hence the bulk dynamics is described by a single transition rate $q$, and the bulk generator is

$$\mathcal{L}_{\text{bulk}}[P_t(\eta)] = \sum_{k=1}^{N-1} q\left[P_t\left(\eta_{\bar{k},\underline{k+1}}\right) + P_t\left(\eta_{\underline{k},\overline{k+1}}\right) - 2P_t(\eta)\right] \tag{S18}$$

eventually, we take $q = 1$.

3. **Local creation/annihilation:** Transitions $\eta \underset{a_k(n_k+1)}{\overset{c_k(n_k)}{\rightleftharpoons}} \eta_{\bar{k}}$ describe the local (in-cell) processes of emission (creation) and absorption (annihilation) of radiation from the walls of cell $k$, independently of all other cells. Building on the analogy with the equilibrium dynamics described in the previous section S III of this Supplemental Material, the corresponding generator is

$$\mathcal{L}^{\perp}[P_t(\eta)] = \sum_{k=1}^{N}\left[c_k(n_k - 1)P_t(\eta_{\underline{k}}) + a_k(n_k + 1)P_t(\eta_{\bar{k}}) - P_t(\eta)\Big(a_k(n_k) + c_k(n_k)\Big)\right] \tag{S19}$$

Altogether, the generator $\mathcal{L}$, of the master equation (5) in the main text becomes

$$\begin{aligned}\mathcal{L}[P_t(\eta)] =& \mathcal{L}^{\parallel}[P_t(\eta)] + \mathcal{L}^{\perp}[P_t(\eta)] \\ =& \alpha\left[P_t(\eta_{\underline{1}}) - P_t(\eta)\right] + \gamma\left[P_t(\eta_{\bar{1}}) - P_t(\eta)\right] \\ &+ \delta\left[P(\eta_{\underline{N}}, t) - P_t(\eta)\right] + \beta\left[P_t(\eta_{\bar{N}}) - P_t(\eta)\right] \\ &+ \sum_{k=1}^{N-1} q\left[P_t\left(\eta_{\bar{k},\underline{k+1}}\right) + P_t\left(\eta_{\underline{k},\overline{k+1}}\right) - 2P_t(\eta)\right] \\ &+ \sum_{k=1}^{N}\Big[c_k(n_k - 1)P_t(\eta_{\underline{k}}) + a_k(n_k + 1)P_t(\eta_{\bar{k}}) \\ &\qquad - P_t(\eta)\Big(a_k(n_k) + c_k(n_k)\Big)\Big].\end{aligned} \tag{S20}$$

**B. Steady-State $P_\infty(\eta)$ and expression of $z_k$**

Here we show that (10) along with (11), from the main text, constitute the steady state for the master equation generated by (S20), i.e we prove (5) of the main text –

$$\mathcal{L}[P_\infty(\eta)] = 0 . \tag{S21}$$

First let us show that the stationary probability

$$P_\infty(\eta) = \prod_{k=1}^{N} (1-z_k) z_k^{n_k} , \tag{S22}$$

is in the kernel of the generator $\mathcal{L}^\perp$. Inserting (S22) into (S19) gives

$$\begin{aligned}
\mathcal{L}^\perp[P_\infty(\eta)] &= \sum_{k=1}^{N} \left[ c_k(n_k-1) \left\{ P_\infty(\eta_{\underline{k}}) - P_\infty(\eta) \frac{a_k(n_k)}{c_k(n_k-1)} \right\} \right. \\
&\quad \left. + a_k(n_k+1) \left\{ P_\infty(\eta_{\overline{k}}) - P_\infty(\eta) \frac{c_k(n_k)}{a_k(n_k+1)} \right\} \right] \\
&= \sum_{k=1}^{N} \left[ c_k(n_k-1) \left\{ P_\infty(\eta_{\underline{k}}) - P_\infty(\eta) \frac{P_\infty(\eta_{\underline{k}})}{P_\infty(\eta)} \right\} \right. \\
&\quad \left. + a_k(n_k+1) \left\{ P_\infty(\eta_{\overline{k}}) - P_\infty(\eta) \frac{P_\infty(\eta_{\overline{k}})}{P_\infty(\eta)} \right\} \right] \\
&= 0 ,
\end{aligned}$$

where we have used the local detailed balance relation (S8) (or (12) in the main text). We now proceed to consider the generator $\mathcal{L}^\parallel$.

Inserting $P_\infty(\eta)$ from (S22), into the kernel equation (S21), leads to

$$\begin{aligned}
0 =& \alpha \left[ P_\infty(\eta_{\underline{1}}) - P_\infty(\eta) \right] + \gamma \left[ P_\infty(\eta_{\overline{1}}) - P_\infty(\eta) \right] \\
&+ \delta \left[ P_\infty(\eta_{\underline{N}}) - P_\infty(\eta) \right] + \beta \left[ P_\infty(\eta_{\overline{N}}) - P_\infty(\eta) \right] \\
&+ \sum_{k=1}^{N-1} q \left( \left[ P_\infty\left(\eta_{\overline{k},\underline{k+1}}\right) - P_\infty(\eta) \right] + \left[ P_\infty\left(\eta_{\underline{k},\overline{k+1}}\right) - P_\infty(\eta) \right] \right) .
\end{aligned}$$

We next use the following property of the stationary probability

$$\begin{aligned}
P_\infty(\eta_{\underline{k}}) &= z_k^{-1} P_\infty(\eta) \\
P_\infty(\eta_{\overline{k}}) &= z_k P_\infty(\eta)
\end{aligned}$$

so that,

$$\begin{aligned}
0 =& \alpha \left[ z_1^{-1} - 1 \right] P_\infty(\eta) + \gamma \left[ z_1 - 1 \right] P_\infty(\eta) \\
&+ \delta \left[ z_N^{-1} - 1 \right] P_\infty(\eta) + \beta \left[ z_N - 1 \right] P_\infty(\eta) \\
&+ \sum_{k=1}^{N-1} q \left( z_k z_{k+1}^{-1} - 1 + z_k^{-1} z_{k+1} - 1 \right) P_\infty(\eta) .
\end{aligned}$$

Since $P_\infty(\eta)$ factors out, we obtain the following relation for $z_k$

$$\begin{aligned}
0 =& \frac{z_1(\gamma z_1 - \alpha) + (\alpha - z_1 \gamma)}{z_1} + \frac{z_N(\beta z_N - \delta) + (\delta - z_N \beta)}{z_N} \\
&+ \sum_{k=1}^{N-1} q \left( \frac{z_k - z_{k+1}}{z_{k+1}} - \frac{z_{k+1} - z_k}{z_k} \right) \\
=& \frac{(\gamma z_1 - \alpha)(z_1 - 1)}{z_1} + \frac{(\beta z_N - \delta)(z_N - 1)}{z_N} \\
&+ \sum_{k=1}^{N-1} q \left( \frac{(z_k - z_{k+1})^2}{z_k z_{k+1}} \right) .
\end{aligned}$$

To satisfy this equation, namely making the right hand side to vanish, we assume

$$q(z_k - z_{k+1}) = \alpha - z_1\gamma = z_N\beta - \delta = c,  \tag{S23}$$

where $c \neq 0$ is a real constant. After factoring out $c$, the equation reads

$$0 = \frac{-z_1 + 1}{z_1} + \frac{z_N - 1}{z_N} + \sum_{k=1}^{N-1} q\left\{\frac{1}{z_{k+1}} - \frac{1}{z_k}\right\}.$$

We observe that the sum in the last term is telescopic,

$$\sum_{k=1}^{N-1}\left[\frac{1}{z_{k+1}} - \frac{1}{z_k}\right] = \frac{1}{z_N} - \frac{1}{z_1},$$

so that the equality holds

$$-1 + \frac{1}{z_1} + 1 - \frac{1}{z_N} + \frac{1}{z_N} - \frac{1}{z_1} = 0$$

We now obtain a closed form solution for $z_k$, as shown in (13) in the main text. First, from $q(z_k - z_{k+1}) = c$ we obtain

$$z_k = -\frac{c}{q}k + Y,$$

where $Y$ is a real constant. It is determined from setting $k = 1$ and comparing to $z_1$ from $\alpha - z_1\gamma = c$,

$$\frac{\alpha - c}{\gamma} = z_1 = -\frac{c}{q} + Y.$$

This gives $Y = c\left(\frac{1}{q} - \frac{1}{\gamma}\right) + \frac{\alpha}{\gamma}$, so that $z_k$ becomes,

$$z_k = -c\left(\frac{k-1}{q} + \frac{1}{\gamma}\right) + \frac{\alpha}{\gamma}.$$

The constant $c$ is determined from setting $k = N$ and comparing to $z_N$ from $z_N\beta - \delta = c$,

$$\frac{c + \delta}{\beta} = z_N = c\left(\frac{1-N}{q} - \frac{1}{\gamma}\right) + \frac{\alpha}{\gamma},$$

which gives

$$c = \frac{\frac{\delta}{\beta} - \frac{\alpha}{\gamma}}{\frac{1-N}{q} - \frac{1}{\gamma} - \frac{1}{\beta}} = \frac{q(\alpha\beta - \gamma\delta)}{\beta\gamma(N-1) + q(\beta + \gamma)}$$

Thus, for $z_k$ we obtain

$$z_k = -c\frac{k}{q} + c\left(\frac{1}{q} - \frac{1}{\gamma}\right) + \frac{\alpha}{\gamma}$$
$$= \frac{c(\gamma(1-k) + q) + q\alpha}{q\gamma}$$
$$= \frac{1}{q\gamma}\frac{q(\alpha\beta - \gamma\delta)(\gamma(1-k) + q) + q\alpha(\beta\gamma(N-1) + q(\beta + \gamma))}{\beta\gamma(N-1) + q(\beta + \gamma)}$$
$$= \frac{k(\gamma\delta - \alpha\beta) - \gamma\delta + \alpha\beta N + q(\alpha + \delta)}{\beta\gamma(N-1) + q(\beta + \gamma)}.$$

From the above, equation (13) from the main text is obtained by setting $q = 1$ and with a simple rearrangement of terms

$$z_k = \frac{\frac{k}{N}(\gamma\delta - \alpha\beta) + \alpha\beta - \frac{1}{N}[\gamma\delta + (\alpha + \delta)]}{\beta\gamma\left(1 - \frac{1}{N}\right) + \frac{1}{N}(\beta + \gamma)}.  \tag{S24}$$

## S V. TRANSPORT COEFFICIENTS

In this section we address an important point regarding the transport coefficients (3) and (20) of the main text. The latter are given in terms of the spectral energy density at the reservoirs, i.e $u_{\mathcal{L}/\mathcal{R}}$, while the former are given in terms of the steady state profile (4), i.e $\bar{u}_\nu(x) = \frac{z_\mathcal{L} + \frac{x}{L}(z_\mathcal{R} - z_\mathcal{L})}{1 - z_\mathcal{L} - \frac{x}{L}(z_\mathcal{R} - z_\mathcal{L})}$. We now show that these two expression are related by continuity of the fugacity $z(x)$ given by (17) in the main text.

First, we observe that expressions (20), in terms of the fugacities at the reservoirs, are (close to equilibrium $u_\mathcal{L} \equiv u$, $u_\mathcal{R} \equiv u + \Delta u$ with $\Delta u \ll 1$)

$$\sigma(u) = 2z(u) \tag{S25a}$$

$$D(u) = \frac{1}{2}\frac{\mathrm{d}\sigma(u)}{\mathrm{d}u}, \tag{S25b}$$

where $z = u/(1+u)$. Therefore, we can rewrite the cumulant generating function (15) from the main text, as

$$\begin{aligned}\mu(\lambda) &= \frac{1}{2L}\left(1 - \mathrm{e}^{-\lambda}\right)\left(\mathrm{e}^\lambda(z' + 2z) + (z' - 2z)\right) \\ &= \frac{2}{L}\sinh\left(\frac{\lambda}{2}\right)\left(D(u)\cosh\left(\frac{\lambda}{2}\right) + \sigma(u)\sinh\left(\frac{\lambda}{2}\right)\right).\end{aligned}$$

From here we proceed as follows: since the fugacity (and its derivative) is continuous both in $u$ and $x$, and the stationary profile (4) obeys the boundary conditions (14), by continuity, the coefficients in (20) are given by (3).

## S VI. TEMPERATURE COMPARISON

In this section we elaborate on the integrals of the spectral entropy (22) and energy (18) densities, and their relation to the temperature function $T_\tau(x)$. Specifically, we define and derive the constants $A_{s/u/T}$.

Starting from the scaling form of the spectral energy density, we have

$$\begin{aligned}u(x) &= \frac{8\pi h}{c^3}\int_0^\infty \mathrm{d}\nu\, \nu^3 \Phi\left(\frac{h\nu}{k_B T_\tau(x)}\right) \\ &= \left\{Y \equiv \frac{h\nu}{k_B T_\tau(x)}\right\} \\ &= \frac{8\pi h}{c^3}\left(\frac{k_B T_\tau(x)}{h}\right)^4 \int_0^\infty \mathrm{d}Y\, Y^3 \Phi(Y) \\ &\equiv A_u T_\tau^4(x)\end{aligned}$$

Hence,

$$A_u = \frac{8\pi k_B^4}{(hc)^3}\int_0^\infty \mathrm{d}Y\, Y^3 \Phi(Y). \tag{S26}$$

For the integral of the spectral entropy density (22), denoting

$$f(\Phi) = -\left[(1+\Phi)\ln(1+\Phi) - \Phi\ln\Phi\right]$$

we have,

$$\begin{aligned}s(x) &= \frac{8\pi k_B}{c^3}\int_0^\infty \mathrm{d}\nu\, \nu^2 f(\Phi) \\ &= \left\{Y \equiv \frac{h\nu}{k_B T_\tau(x)}\right\} \\ &= \frac{8\pi k_B}{c^3}\left(\frac{k_B T_\tau(x)}{h}\right)^3 \int_0^\infty \mathrm{d}Y\, Y^2 f(\Phi(Y)) \\ &\equiv A_s T_\tau^3(x)\end{aligned}$$

so that,

$$A_s = \frac{8\pi k_B^4}{(hc)^3} \int_0^\infty dY\, Y^2 f\left(\Phi\left(Y\right)\right). \tag{S27}$$

It is worth noting that since the scaling function $\Phi$ is not the Planck function, i.e

$$\Phi\left(Y\right) \neq \frac{1}{e^Y - 1},$$

the ratio of the constants depends only on the integrals

$$\frac{A_s}{A_u} = \frac{\int_0^\infty dY\, Y^2 f\left(\Phi\left(Y\right)\right)}{\int_0^\infty dY\, Y^3 \Phi\left(Y\right)}. \tag{S28}$$

Hence the derivative of the entropy density with respect to the energy density is

$$\begin{aligned}\frac{\partial s\left(x\right)}{\partial u\left(x\right)} &= \frac{\partial}{\partial u\left(x\right)} A_s T_\tau^3\left(x\right) \\ &= A_s \frac{\partial}{\partial u\left(x\right)} \left(\frac{u\left(x\right)}{A_u}\right)^{\frac{3}{4}} \\ &= \frac{A_s}{A_u^{3/4}} \frac{3}{4} u^{-\frac{1}{4}}\left(x\right) \\ &= \frac{A_s}{A_u^{3/4}} \frac{3}{4} \frac{1}{A_u^{1/4}} \frac{1}{T_\tau\left(x\right)} \\ &= \frac{3}{4} \frac{A_s}{A_u} \frac{1}{T_\tau\left(x\right)}.\end{aligned}$$

Thus, we deduce the last constant

$$A_T = \frac{3}{4} \frac{\int_0^\infty dY\, Y^2 f\left(\Phi\left(Y\right)\right)}{\int_0^\infty dY\, Y^3 \Phi\left(Y\right)}. \tag{S29}$$

At equilibrium, $\tau = 1$, the scaling function $\Phi$ is the Planck function and the ratio of integrals (S28) equals $4/3$ thus setting $A_{T,\mathrm{eq}} = 1$.

---

[S1] L. Bertini, A. de Sole, D. Gabrielli, G. Jona-Lasinio, and C. Landim, Journal of Statistical Physics **107**, 635 (2002).
[S2] C. Kipnis and C. Landim, *Scaling limits of interacting particle systems*, Vol. 320 (Springer Science & Business Media, 1998).
[S3] According to the Frobenius-Perron theorem, there must be one, and only one, stationary state for this dynamics.


\fi

\end{document}